\documentclass[preprint]{aastex5b}
\usepackage{psfig,natbib}
\received{24 January 2001}  
\accepted{21 February 2001}

\newcommand{\citepeg}[1]{\citep[{e.g.,}][]{#1}}
\newcommand{\citepcf}[1]{\citep[{cf.}\phantom{}][]{#1}}

\def\lsim{\hbox{ \rlap{\raise 0.425ex\hbox{$<$}}\lower 0.65ex\hbox{$\sim$}}}
\def\gsim{\hbox{ \rlap{\raise 0.425ex\hbox{$>$}}\lower 0.65ex\hbox{$\sim$}}}

\def\ale{\mathrel{\hbox{\rlap{\hbox{\lower4pt\hbox{$\sim$}}}\hbox{$<$}}}}
\def\age{\mathrel{\hbox{\rlap{\hbox{\lower4pt\hbox{$\sim$}}}\hbox{$>$}}}}

\slugcomment{Accepted to The Astronomical Journal}
\shorttitle{The Prompt Energy Release of GRBs} \shortauthors{Bloom,
Frail, \& Sari}
\begin{document}
\title{The Prompt Energy Release of Gamma-Ray Bursts using a cosmological $k$-correction}

\author{Joshua S.~Bloom}
\affil{Palomar Observatory 105--24, California Institute of
       Technology, Pasadena, CA 91125, USA}
\author{Dale A.~Frail}
\affil{National Radio Astronomy Observatory, Socorro, NM 87801, USA}
\author{Re'em Sari}
\affil{Theoretical Astrophysics 130-33, California Institute of
       Technology , Pasadena CA 91125, USA}
%
\begin{abstract}

The fluences of gamma-ray bursts (GRBs) are measured with a variety of
instruments in different detector energy ranges.  A detailed
comparison of the implied energy releases of the GRB sample requires,
then, an accurate accounting of this diversity in fluence measurements
which properly corrects for the redshifting of GRB spectra. Here, we
develop a methodology to ``$k$-correct'' the implied prompt energy
release of a GRB to a fixed co-moving bandpass.  This allows us to
homogenize the prompt energy release of 17 cosmological GRBs (using
published redshifts, fluences, and spectra) to two common co-moving
bandpasses: 20--2000 keV and 0.1 keV--10 MeV (``bolometric'').  While
the overall distribution of GRB energy releases does not change
significantly by using a $k$-correction, we show that uncorrected
energy estimates systematically undercounts the bolometric energy by
$\sim$5\% to 600\%, depending on the particular GRB. We find that the
median bolometric isotropic-equivalent prompt energy release is
\hbox{2.19 $\times 10^{53}$ erg} with an r.m.s.~scatter of 0.80 dex.
The typical estimated uncertainty on a given $k$-corrected energy
measurement is $\sim$20\%.
\end{abstract}

\section{Introduction}

Four years into the era of gamma-ray burst (GRB) afterglow studies,
there are currently 17 GRBs with measured redshifts.  Arguably the
most important utility of GRB redshift measurements is the opportunity
to determine the total energy release.  The total energy release is,
of course, a critical diagnostic of the progenitors of GRBs in that it
affords a direct probe of the energy (and hence mass) reservoir in the
progenitor system. In fact, an estimate of the isotropic energy
radiated from GRB 971214 of $3 \times 10^{53}$ erg (about 10\% the
restmass energy of a neutron star) placed uncomfortable constraints on
most viable stellar mass progenitors \citep{kdr+98}. Not all GRBs have
estimated energies as high as GRB 971214; indeed, estimated GRB
energies range over nearly four orders of magnitude and appear to be
anything but standard candles.  This fact, in and of itself, may point
to a diversity of progenitor scenarios and/or emission geometries of
GRBs as a class.

Well before the first redshift determination of GRBs, considerable
effort was devoted towards divining the total prompt energy release
(or peak luminosity) in GRBs by examining the brightness distribution
of GRB events \citep{pir92,feh+93,wl94,fb95,hmh+96,mpp96}.  The
so-called ``log $N$-log $S$'' or ``log $N$-log $P$'' distributions,
where $S$ ($P$) is fluence (peak flux) and $N$ is the number of bursts
observed above that fluence (peak flux), was known to exhibit a
roll-over at faint fluences and fluxes \citep{mfw+92}. This paucity of
faint events was believed to have arisen from cosmological expansion
effects if the faintest bursts originated beyond a redshift $z \sim
1$.  Two major assumptions were required in these studies to extract a
meaningful energy scale: first, that GRBs are standard candles in peak
flux or energy; and second, that the trigger efficiencies for faint
events were known well-enough to correct the observed brightness
distributions.  The first assumption was dramatically disproved after
redshifts were found for GRB 970508 \citep{mdk+97} and GRB 971214
\citep{kdr+98}, where the estimated isotropic energy release was 7
$\times 10^{51}$ erg and 3 $\times 10^{53}$ erg,
respectively. \citet{sch99} and \citet{klk+00} have recently
reexamined the log $N$-log $P$ distributions in the context of known
GRB redshifts and relax the standard candle assumption.

The measurement of a fluence/flux of a given GRB is limited by the
sensitivity range (``bandpass'') of the detector; thereby, to
ascertain the total (``bolometric'') fluence/flux one needs to
extrapolate the observed spectrum outside the detector
bandpass. Furthermore, the same GRB placed at different cosmological
distances would, even after accounting for the 1/$R^2$ dimming, result
in different fluence measurements since its co-moving spectrum would
be redshifted.  In the extreme case, a GRB originating from a very
high redshift might only be detectable in the X-ray
bandpass. Therefore, in order to determine the energy of a set of GRBs
in some common co-moving bandpass, it is not enough to measure the
brightness distribution in a common detector bandpass; instead, one
must use the spectra of the GRBs themselves to correct for the
redshifting effect.

By analogy with a photometric technique often employed in
observational cosmology, we call this correction a
$k$-correction. Before the first redshifts of GRBs were known,
\citet{feh+93}, using the standard candle assumption and the log
$N$-log $P$ distribution, were the first to determine peak fluxes in a
common co-moving bandpass using a $k$-correction.  Later, \citet{fb95}
used a similar technique using more realistic GRB spectra to find the
standard candle peak fluxes.  \citet{bfi96} extended the
$k$-correction technique to examine the energy scale implied from the
log $N$-log $S$ distribution.

Clearly, to begin to understand and model the prompt energy release
distribution, we need to determine GRB energies in a common co-moving
bandpass.  In this paper, we find the $k$-corrected energies of GRBs
with known redshifts in order to homogenize the set of GRB energies to
a common co-moving bandpass. The correction methodology, described in
\S \ref{sec:method}, is straightforward when information on the GRB
spectrum is known.  However, a number of GRBs have only a fluence and
a redshift published; to estimate the $k$-correction, then, we use an
ensemble of template GRB spectra, expanding upon the earlier work of
\citet{bfi96}.  In \S \ref{sec:results} we present the $k$-corrected
energies for 17 cosmological GRBs (plus 6 additional well-studied GRBs
with an assumed redshift) in both a co-moving bandpass of 20--2000 keV
and a bolometric bandpass. We employ a number of tests to show that
$k$-corrections derived from the template spectra method are robust.
Lastly, we point out that by assuming that the total prompt energy
release is simply $E = 4 \pi D_l^2 S_{\rm obs}/(1+z)$, where $S_{\rm
obs}$ is the quoted fluence in some detector bandpass, undercounts the
bolometric energy release by 5\% to 600\%, depending on the particular
GRB.

\section{Methodology}
\label{sec:method}

The relationship of the bolometric energy ($E_{\rm bol}$) to the
bolometric fluence ($S_{\rm bol}$) is
\begin{equation}
E_{\rm bol} = \frac{4 \pi D_l^2}{1+z}\, S_{\rm bol},
\label{eq:edef}
\end{equation}
where $D_l$ is the luminosity distance to the source at redshift $z$.
The bolometric fluence is difficult (if not impossible) to measure
directly. Instead fluence (energy per unit area) is typically measured
in some detector bandpass bracketed by the energies, $e_1$ and $e_2$.
The Burst and Transient Source Experiment (BATSE), for instance, was
capable of measuring fluence in the bandpass $e_1 = 20$ keV to $e_2 =
2000$ keV. We now define this bandpass fluence, $S_{\rm obs} \equiv
S_{[e_1,e_2]}$, as
\begin{equation}
S_{[e_1,e_2]} = \int_{e_1}^{e_2} E\, S_0 \phi(E) dE,
\end{equation}
where the (time-integrated) spectral shape of the GRB, $\phi(E)$, and
the normalization $S_0$ (units of photon per unit area per unit
energy) are found from measurements. In general, however, we wish to
measure the energy, $E_{[E_1, E_2]}$, in some fixed co-moving bandpass,
bracketed by two arbitrary energies $E_1$ and $E_2$. In this case,
\begin{equation}
E_{[E_1, E_2]} = \frac{4 \pi D_l^2}{1+z}\, S_{[\frac{E_1}{1+z},\frac{E_2}{1+z}]}
\label{eq:egen}
\end{equation}
Note that by letting $E_1 \rightarrow 0$ and $E_2 \rightarrow \infty$
we recover the bolometric result in equation \ref{eq:edef}.  In
reality, we can only measure, $d E_{[E_1, E_2]}/d \Omega$, the prompt
energy release per unit solid angle in the direction of the GRB
detector. The energy derived using equation \ref{eq:egen}, then, is
the ``isotropic-equivalent'' energy release.  That is,
$E_{[E_1, E_2]} = 4 \pi\, d E_{[E_1, E_2]}/d \Omega$ is the actual
energy release if the GRB radiates isotropically.

We rewrite equation \ref{eq:egen} in terms of $S_{[e_1,e_2]}$
so that,
\begin{equation}
E_{[E_1, E_2]} = S_{[e_1,e_2]} \frac{4 \pi D_l^2}{1+z}\, k[e_1, e_2,
E_1, E_2, z, \phi(E)],
\label{eq:theken}
\end{equation}
where $k[e_1,e_2,E_1,E_2,z]$ is what we take as the cosmological
$k$-correction,
\begin{equation}
k = k[e_1,e_2,E_1,E_2,z, \phi(E)] \equiv
\frac{I{[\frac{E_1}{1+z},\frac{E_2}{1+z}]}}{I{[e_1,e_2]}},
\end{equation}
with $I [x_1, x_2] = S_{[x_1, x_2]} / S_0$. Note that $k = 1$ when
$E_1 = e_1 (1 + z)$ and $E_2 = e_2 (1 + z)$.  That is, there is no
$k$-correction when the fixed co-moving bandpass corresponds precisely
to the redshifted detector bandpass. To perform the above calculation,
the spectrum must be generally known outside the bandpass energy range
where the fluence was measured. Since spectral data on a given GRB are
typically sparse, it is common practice to fit $\phi(E)$ analytically
using a broken power-law functional form suggested by \citet{bmf+93}.
The Band et al.~spectral shape is given by,
\begin{eqnarray}
\phi(E) & = & \left(\frac{E}{{\rm 100~keV}}\right)^{\alpha}\, 
	       \exp \left(- \frac{E}{E_0} \right), 
            ~~~~~~~~~~~~~~~~~~~~~~~~~~~~ (\alpha - \beta)E_0 \ge E,
	       \label{eq:band}\\
        & = & \left[\frac{(\alpha - \beta)\, E_0}
	                 {\rm {100~keV}}\right]^{(\alpha - \beta)}
              \exp (\beta - \alpha)\left(\frac{E}{{\rm 100~keV}}\right)^\beta,
            ~~~~~~~~~~(\alpha - \beta)E_0 \le E \nonumber,
\end{eqnarray}
This three parameter analytic fit to the spectral shape (along with
the normalization $S_0$) appears to adequately describe GRB spectra in
the bandpass energy range of BATSE \citep[20--2000
keV;][]{bmf+93,ppb+98} and extended out to tens of MeV \citep{tav96}.

\subsection{Uncertainty Estimates}

In practice, in order to determine the fluence of a GRB, a spectrum is
assumed and forward-folded through the detector response to produce a
predicted total count for that particular detector \citep[see][for a
review]{pbm+00}.  The shape and normalization of the spectrum is then
found simultaneously by iteration until the predicted counts best
matches the raw counts.  As such, there is likely to be some level of
covariance between $S_0$ and the parameters of $\phi(E)$.  However,
probably owing to the complex dependence upon the detector responses,
these covariances are rarely reported and so, in the subsequent error
estimate, we make the simplifying assumption that $\phi(E)$ and the
spectrum normalization $S_0$ are uncorrelated.  (We have tested that
if the parameters are maximally correlated, then the uncertainty
estimate below using the uncorrelated assumption is still reasonably
accurate to a factor of $\sim$2).

The fractional uncertainty in GRB redshift measurements (and hence
$D_l$) are negligible compared to the fractional uncertainty in the
fluence measurement and we therefore ignore this contribution in the
error analysis. The uncertainty, $\sigma_{E_{[E_1,E_2]}}$, in the
energy determination then is given by,
\begin{equation}
\sigma_{E_{[E_1, E_2]}}^2 = 
\left(\frac{ E_{[E_1, E_2]}}{S_{[e_1,e_2]}}\right)^2\sigma^2_{S_{[e_1,e_2]}}
+\left(\frac{ E_{[E_1, E_2]}}{k}\right)^2\sigma^2_{k}.
\label{eq:theerr}
\end{equation}
Adopting the functional form for $\phi(E)$ in equation \ref{eq:band}
and again assuming that $\alpha$, $\beta$, and $E_0$ are uncorrelated,
\begin{eqnarray}
\sigma^2_{k} & = &
  \left(\frac{k}{ I{[\frac{E_1}{1+z},\frac{E_2}{1+z}]}}
	\frac{\partial I{[\frac{E_1}{1+z},\frac{E_2}{1+z}]}}{\partial \alpha}
      - \frac{k}{I{[{e_1},{e_2}]}}
        \frac{\partial I{[{e_1},{e_2}]}}{\partial \alpha}
 \right)^2      \sigma^2_\alpha \\
             & & + 
\left(\frac{k}{ I{[\frac{E_1}{1+z},\frac{E_2}{1+z}]}}
	\frac{\partial I{[\frac{E_1}{1+z},\frac{E_2}{1+z}]}}{\partial \beta}
      - \frac{k}{I{[{e_1},{e_2}]}}
        \frac{\partial I{[{e_1},{e_2}]}}{\partial \beta}
 \right)^2      \sigma^2_\beta \nonumber \\
             & & + 
\left(\frac{k}{ I{[\frac{E_1}{1+z},\frac{E_2}{1+z}]}}
	\frac{\partial I{[\frac{E_1}{1+z},\frac{E_2}{1+z}]}}{\partial E_0}
      - \frac{k}{I{[{e_1},{e_2}]}}
        \frac{\partial I{[{e_1},{e_2}]}}{\partial E_0}
 \right)^2      \sigma^2_{E_0} \nonumber
\end{eqnarray}
with,
\begin{eqnarray}
\frac{\partial I[{x_1},{x_2}]}{\partial \alpha}
  & = & \int_{x_1}^{x_2} E \phi(E) \ln \left(\frac{E}{{\rm
  100~keV}}\right) dE ~~~ \left[=
  \int_{x_1}^{x_2} E \phi(E) \ln \left(\frac{(\alpha - \beta)\,E_0}{{\rm
  100~keV}}\right) dE \right] \label{eq:banderr}
\\
\frac{\partial I[{x_1},{x_2}]}{\partial \beta}
  & = & 0 ~~~~~~~~~~~~~~~~~~~~~~~~~~~~~~~~~~~~~~ \left[=
   \int_{x_1}^{x_2} E \phi(E) \ln \left(\frac{E}{(\alpha -
  \beta)\,E_0}\right) dE \right] \nonumber \\
\frac{\partial I[{x_1},{x_2}]}{\partial E_0}
  & = &  \int_{x_1}^{x_2} \phi(E) \frac{E^2}{E_0^2} dE ~~~~~~~~~~~~~~~~~~ \left[=
    \int_{x_1}^{x_2} \phi(E) \frac{(\alpha - \beta)\, E}{E_0} dE
  \right] 
      \nonumber \\
  & & ~~~~~~~~~~~~~~~~~~~~~~~~~~~~~~~~~~~~~~~(\alpha - \beta)E_0 \ge E
  ~~~ [(\alpha - \beta)E_0 \le E] \nonumber
\end{eqnarray}
Since we often need to extrapolate the spectral shape beyond where the
spectrum is observed, systematic uncertainties may arise when a
particular GRB spectrum is not well-fit by the Band spectral shape
outside the observed bandpass (such as if there are spectral features
in hard X-rays or MeV energies).  However, since most of the energy of
a GRB is emitted near $E_0$ (typically 100--400 keV in the Band et
al.~1993 sample), we do not expect that the spectrum extrapolation
will grossly effect the derived energy.

\subsection{Template Spectra Method}

Often, only the fluence measure is given and no information about the
GRB spectrum is provided. When no spectral information is provided in
a given reference, we need to make an assumption about the spectrum in
order to carry through the analysis above.  Since the spectra of GRBs
are rather diverse in energy cut-off and spectral indices
\citepeg{bmf+93}, it is not necessarily sufficient to assume that the
spectrum of all GRBs are the same, nor are they simple-power laws.

Instead, when no spectral information is provided, we assumed that
$\phi(E)$ was each of the 54 ``template'' spectra fit in
\citet{bmf+93} and then found the implied isotropic energy release
using the GRB redshift and fluence measurement.  We take as the
estimated energy 1.06 $\times$ the median energy implied over the
ensemble of spectra. This small numerical correction factor was found
empirically by comparing the template spectra $k$-corrections to the
$k$-corrections found using the observed spectral data (see figure
\ref{fig:encomp}). For the uncertainty in the energy estimate, we add
in quadrature the statistical error on the fluence normalization (term
1 of equation \ref{eq:theerr}) and the r.m.s.~of the energies implied
using the ensemble of 54 template spectra after twice cleaning for
3-$\sigma$ outliers. As we will demonstrate below (\S
\ref{sec:templ}), using the template spectra method reasonably
recovers the ``true'' energy of those bursts for which a spectrum is
known.

\section{Results}
\label{sec:results}

Tables \ref{tab:allspect1} and \ref{tab:allspect2} give the results of
the $k$-corrected energies in the common energy range 20 to 2000 keV
and 0.1 keV to 10 MeV (``bolometric''), respectively, for all
published fluence and spectra of 23 GRBs with well-studied afterglow.
We assume a cosmology with $H_0 = 65$ km s$^{-1}$ Mpc$^{-1}$,
$\Omega_m = 0.3$, and $\Omega_\Lambda$ = 0.7. The redshifts of 17 of
these GRBs are known.  In the cases where no GRB redshift has been
found (GRB 980329, GRB 980519, GRB 981220, GRB 981226) we assume $z =
1.5$, excepting GRB 980326 where we assume $z=1.0$ \citepcf{bkd+99}.
For completeness, we include GRB 980425 in the tables, though its
association with SN 1998bw (and hence the implied extremely low
redshift) is still in question; the subsequent analysis excludes GRB
980425.

In most cases the average fitted parameters to a Band et al.~spectral
shape was provided in the respective reference. Where the
uncertainties in $S_{\rm obs}$, $\alpha$, $\beta$, and $E_0$ are
provided, we find the uncertainty in the derived energy using
eqns.~(\ref{eq:theerr})---(\ref{eq:banderr}). When no uncertainties
were given we assumed (conservatively) 10\% errors in $S_{\rm obs}$
and $E_0$ and an error of 0.2 in the parameters $\alpha$ and
$\beta$. When no spectral information is given, we use the template
spectral method described above to estimate the $k$-corrected energy
release.  We show the $k$-corrected energies of 17 cosmological GRBs
with confirmed redshifts (not including GRB 980425) in figure
\ref{fig:en-v-z}.

\subsection{Testing the Energy Corrections}
\label{sec:templ}

We now demonstrate the robustness of the energy $k$-corrections and
the adequacy of the template spectral method with several
tests. First, in many cases the same GRB was measured by several
instruments allowing for an intercomparison of implied energies for a
given GRB. In 22 out of 25 intercomparisons in table
\ref{tab:allspect1}, the $k$-corrected energies agree to within
2-$\sigma$ for the same GRB. Second, in figure \ref{fig:encomp}, we
compare the energies of those bursts where spectral information is
provided with the energy implied by assuming that no spectral
information is provided and instead using the template spectra method
to determine the energy.  The comparison between the energies derived
from the actual spectra and those estimated from the template spectra
method show good agreement within the statistical
uncertainties. Third, as illustrated in figure \ref{fig:encomp}, there
are no apparent systematic discrepancies with redshift, $k$-corrected
energies, or co-moving bandpass choice.  Furthermore, we see no
apparent systematic discrepancies with the choice of different
cosmologies.  This strengthens our claim that the $k$-corrected
energies can be robustly determined even without spectral information
for a given GRB.

Implicit in the energy estimate of bursts without published spectral
information is the assumption that the template spectra are from
representative bursts at the same redshift as the GRB in question.
Previously, under the standard candle hypothesis, the template spectra
were first blueshifted back to the presumed redshift of the particular
template burst, then redshifted to the distance of GRB in question
\citep{fb95,bfi96}.  The reason for this was that the template spectra
tended to be of brighter bursts ({\it i.e.}~lower-redshift bursts) and
therefore on the whole, by anzatz, spectrally harder than bursts at
higher redshifts.  Empirically, as seen in figure \ref{fig:encomp}, we
have shown that no such redshift-dependent correction is required
since the template spectra adequately reproduce the implied energies
of those bursts where the spectrum is observationally determined. This
implies that bright (peak flux) GRBs originate from a large range in
redshift, rather than from systematically small redshifts, in
agreement with what is already known from GRB redshift measurements.

In tables \ref{tab:ensum1} and \ref{tab:ensum2}, we present the
implied $k$-corrected prompt GRB energy releases assuming three
different cosmological world models.  Where multiple fluence/spectral
references are available for a given GRB we choose the reference with
the most detailed analysis where we are most confident in the
result. We emphasize again, however, that the $k$-corrected energies
of 22 out of 25 multiple-reference GRBs are self-consistent, as they
should, to within the uncertainties.

\section{Discussion and Summary}

\begin{figure*}
\centerline{\psfig{file=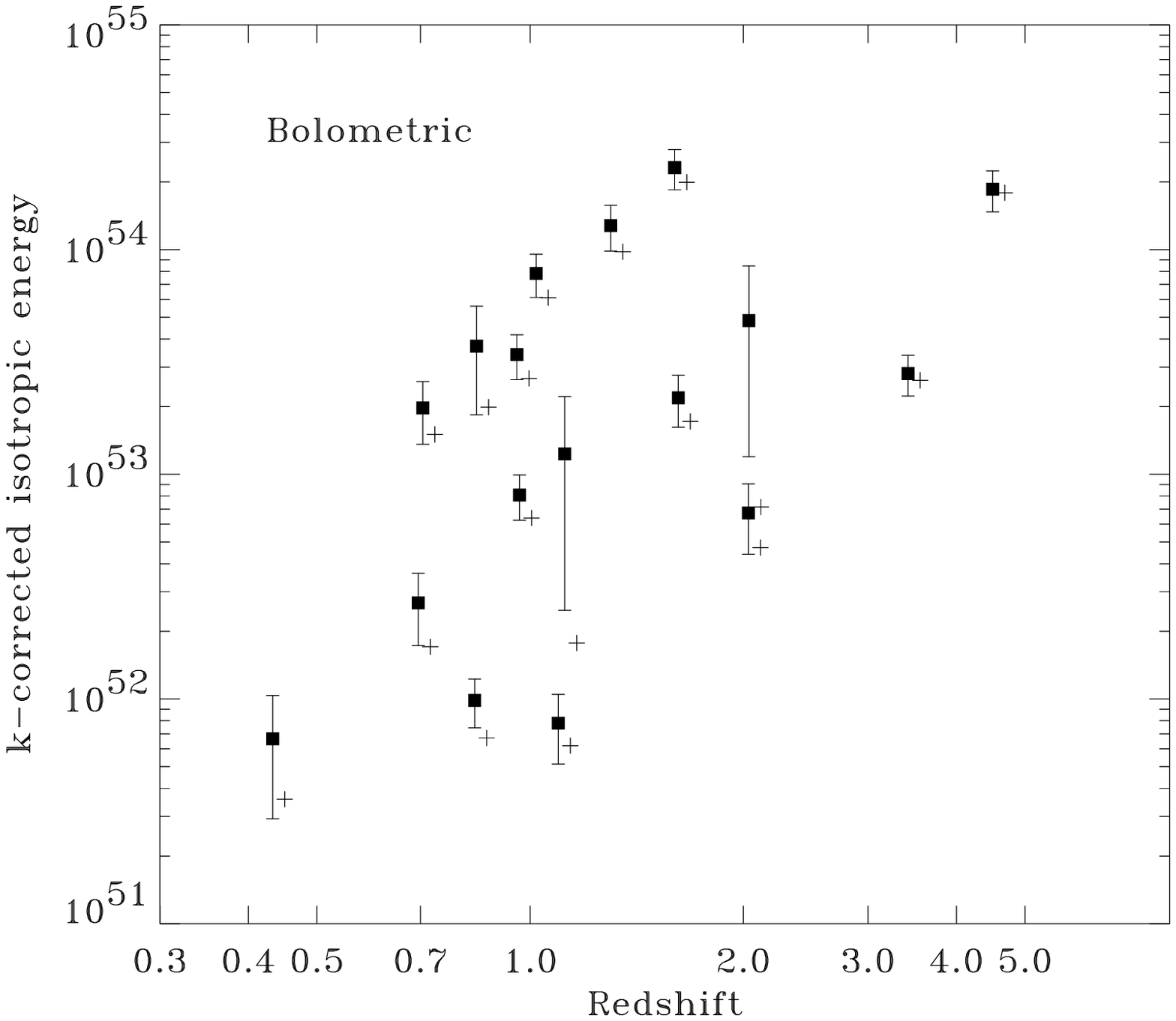,width=3.7in}
	    \psfig{file=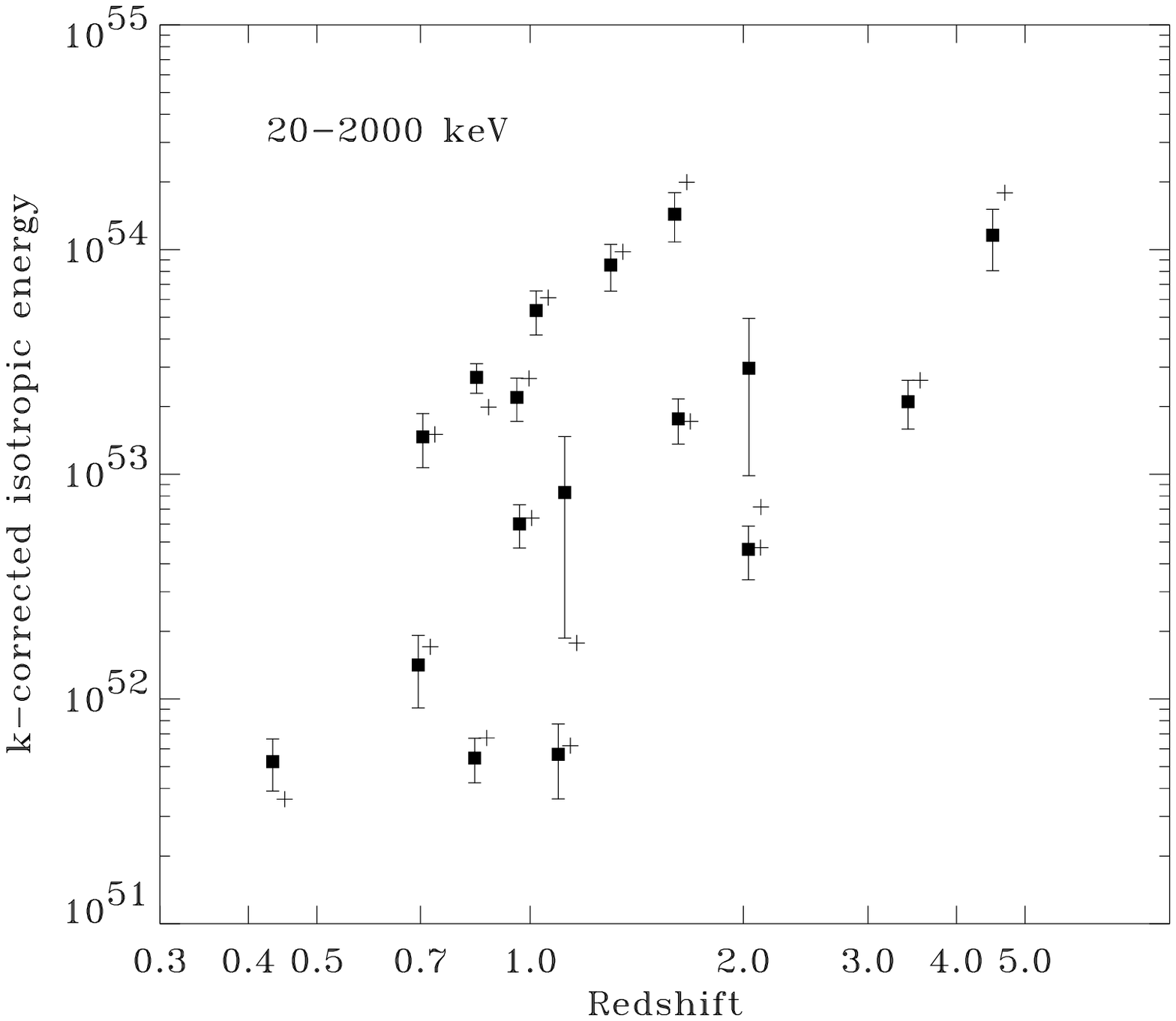,width=3.7in}}
\centerline{\psfig{file=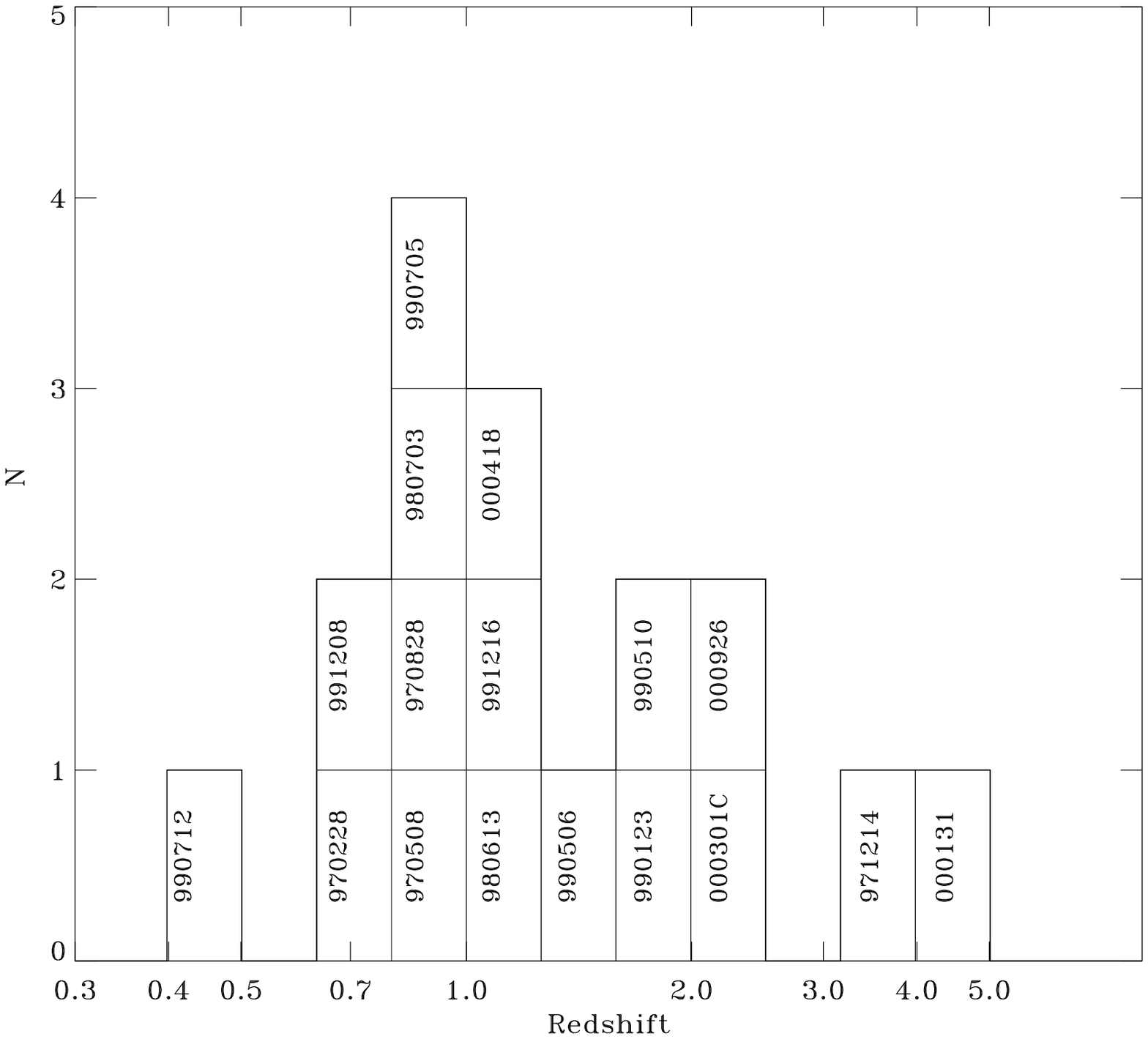,width=3.3in}}
\caption[]{(top) The energy versus redshift relationship for
$k$-corrected bolometric (left) and co-moving 20--2000 keV (right)
prompt energy releases of 17 cosmological GRBs.  The $k$-corrected
energies are noted with a filled square and the errors bars are the
estimated 2 $\sigma$ uncertainties.  The crosses (offset in redshift
for clarity) are the implied energies if no $k$-correction is employed
(that is, simply assuming that $E = 4 \pi D_l^2 S_{\rm obs}/1+z$
). (bottom) The histogram of GRB redshifts binned with equal log
spacing.}
\label{fig:en-v-z}
\end{figure*}

\begin{figure*}
\centerline{\psfig{file=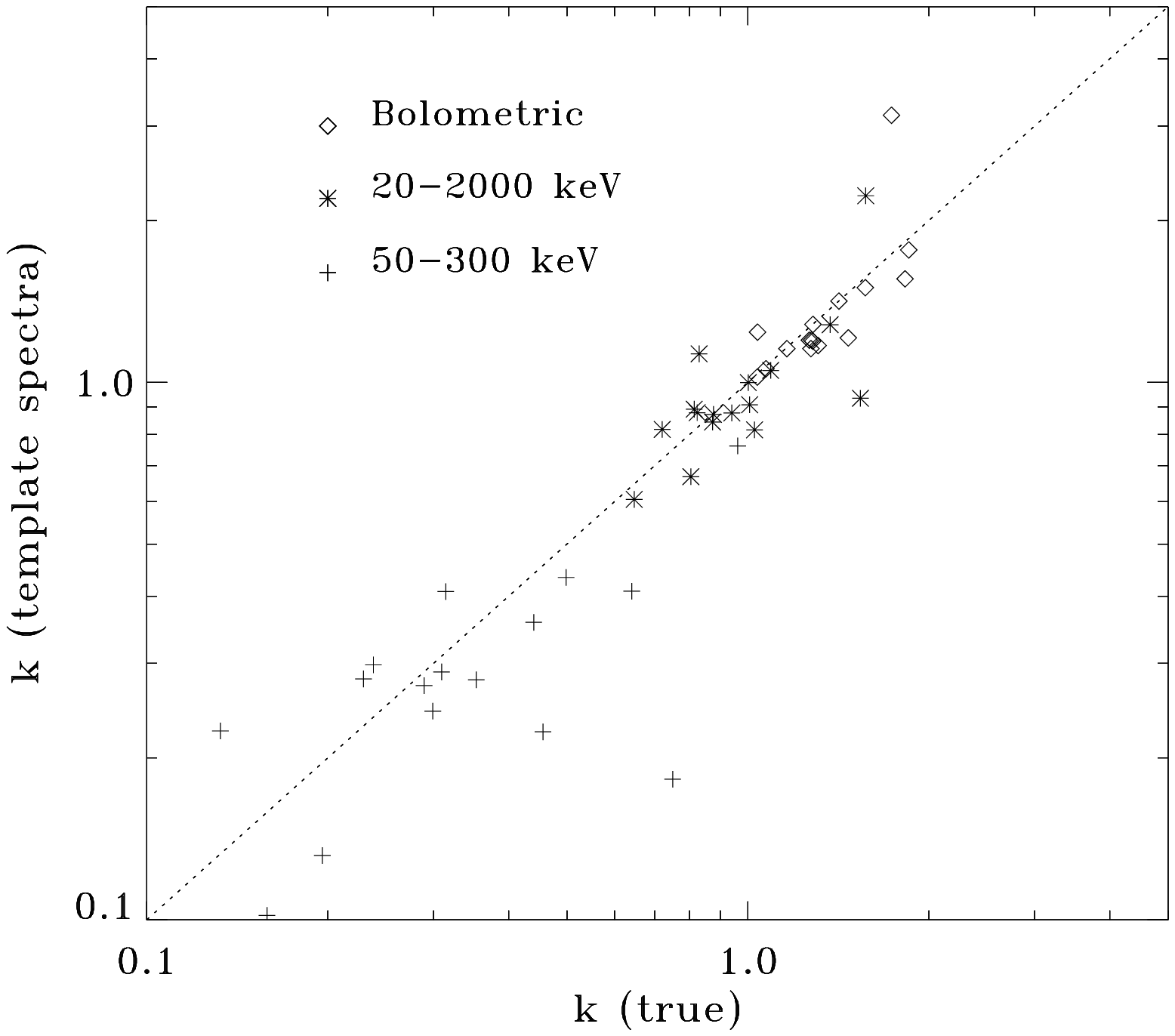,width=4.5in}}
\centerline{\psfig{file=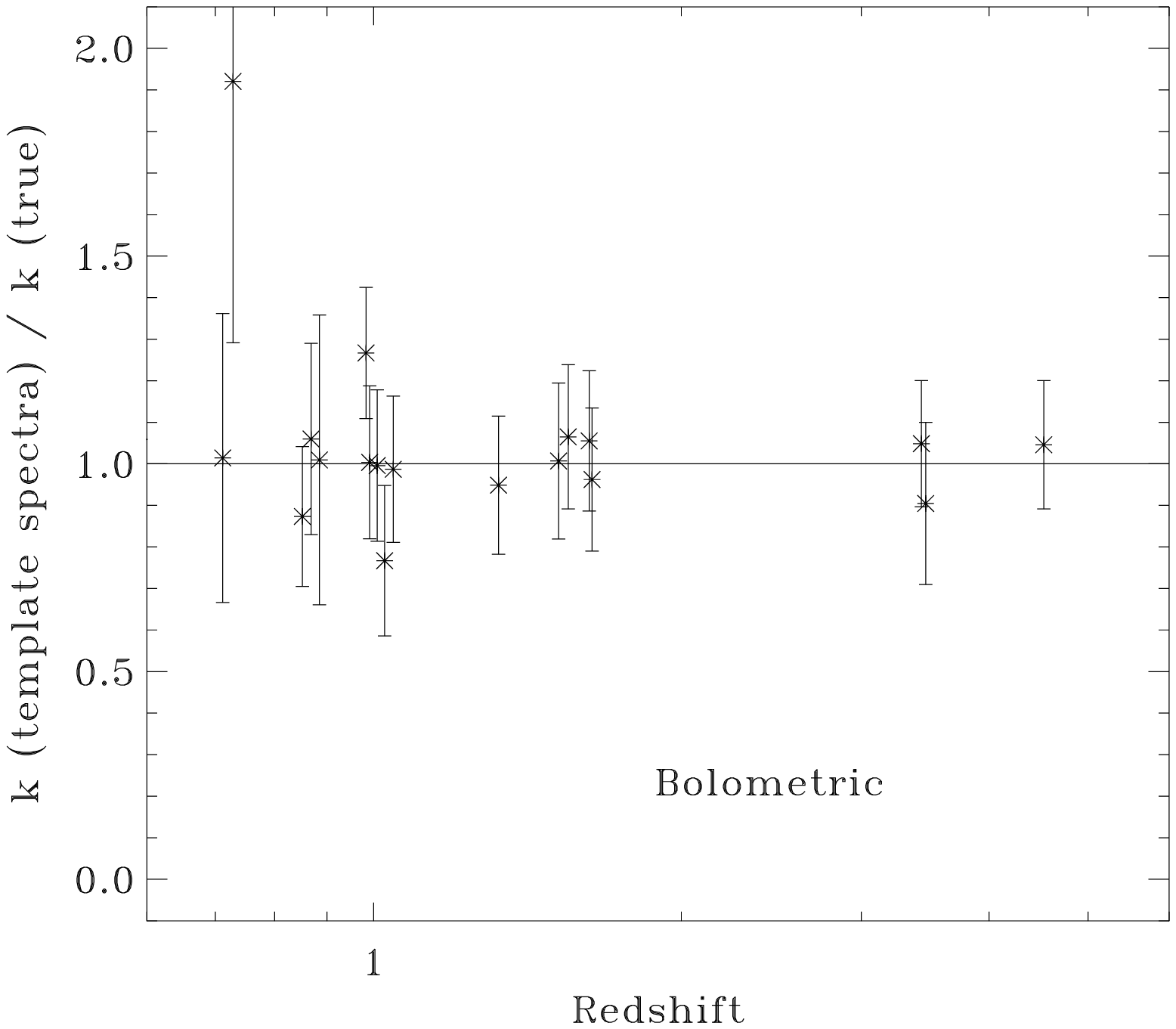,width=3.3in}
	    \psfig{file=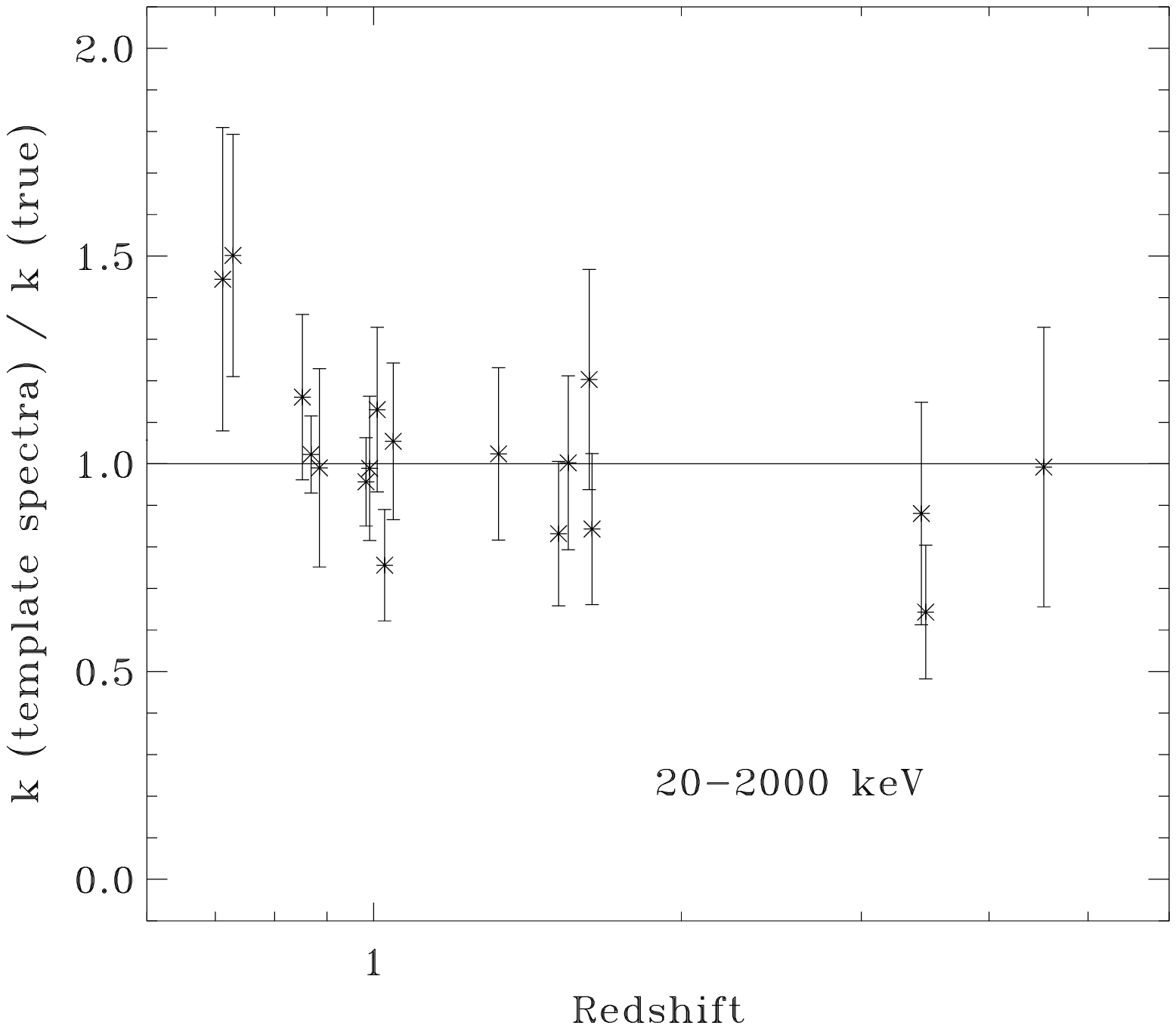,width=3.3in}}
\caption[]{Testing the robustness of the $k$-corrected prompt energy
release derived using the template spectra method described in the
text.  (top) The comparison of the true $k$-correction factor of those
GRBs with both a known redshift and an observed spectrum to the
$k$-correction factor estimated using template spectra method
described in the text.  Though there is some scatter about the
(dashed) line of equal value, there are no apparent systematic
problems in the template spectra method with choice of co-moving
bandpass nor $k$-correction factors.  The figures at bottom show the
ratio of the $k$-corrected GRB energies using the spectral fits to
$k$-corrected energy estimated using the 54 \citet{bmf+93} template
spectra versus GRB redshift.  The redshifts of a given GRB with
multiple spectrum references have been slightly spaced for clarity.
There is no significant correlation between the ratio and redshift.}
\label{fig:encomp}
\end{figure*}

We have found the $k$-corrected isotropic-equivalent prompt energy
release and the associated uncertainties for 17 GRBs with redshifts
and 6 additional GRBs with measured afterglows where we assume a
redshift. Our method for determining the $k$-correction when no
spectral information is known for a particular GRBs appears robust
(see figure~\ref{fig:encomp}). As can be seen in the ``$k$'' column in
table \ref{tab:ensum2}, the simple assumption that $E = 4 \pi D_l^2
S_{\rm obs}/(1+z)$, where $S_{\rm obs}$ is the observed fluence,
systematically undercounts the bolometric energy release by $\sim$5\%
to 600\%, depending on the particular GRB. These differences between
the $k$-corrected energies and the simple uncorrected energies are
highlighted in figure \ref{fig:en-v-z} where the crosses depict the
simple uncorrected energies.  Though in many cases the two energies
are similar, in some the differences are quite large (particularly
those GRBs where fluences were measured in a small bandpass energy
range).  While most $k$-corrections are of order unity, the GRB energy
distribution spans nearly 3 orders of magnitude. As such, the
distribution of uncorrected energies is qualitatively similar to the
distribution of $k$-corrected energies.

\begin{figure*}
\centerline{\psfig{file=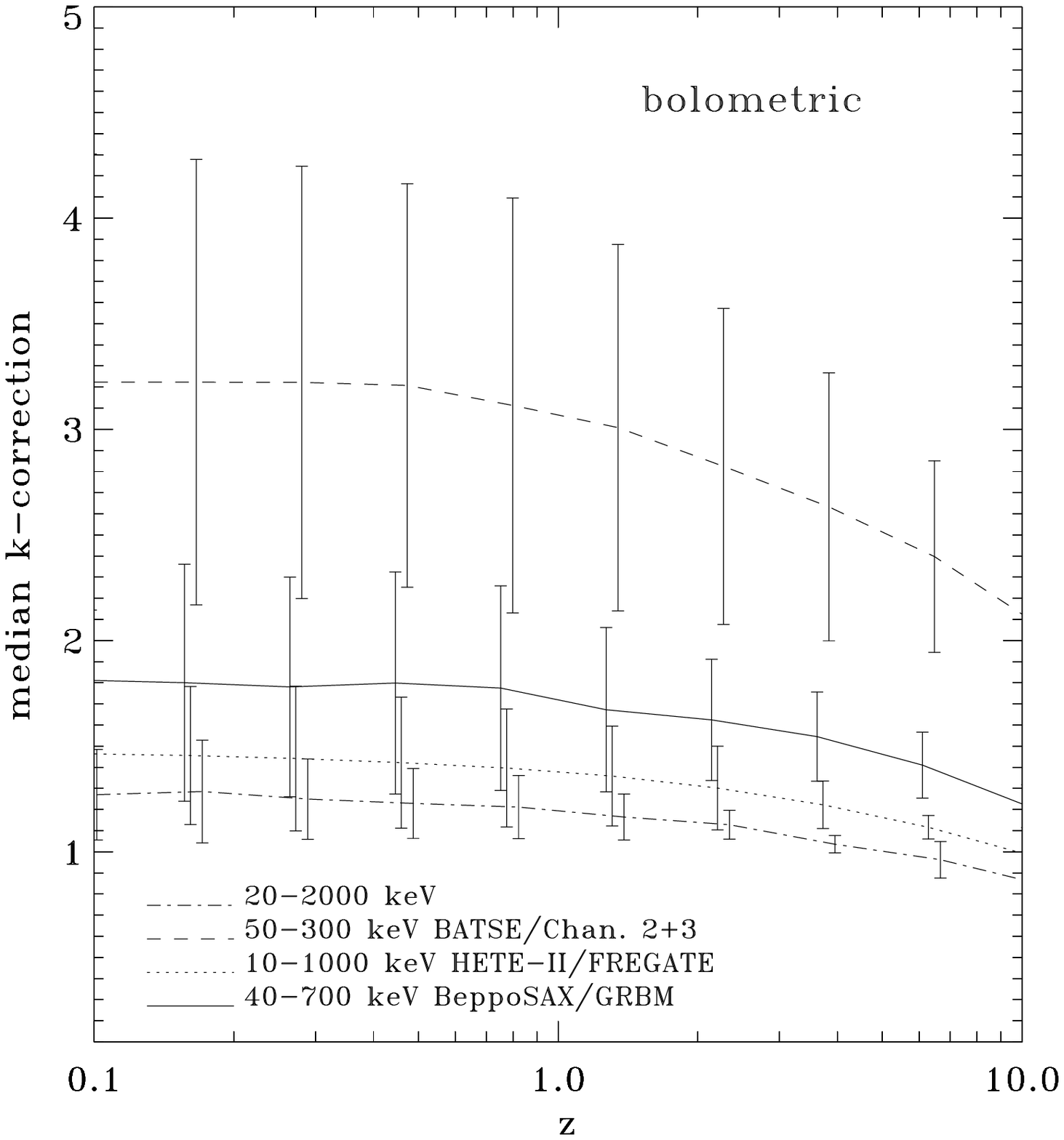,width=3.5in}
           \psfig{file=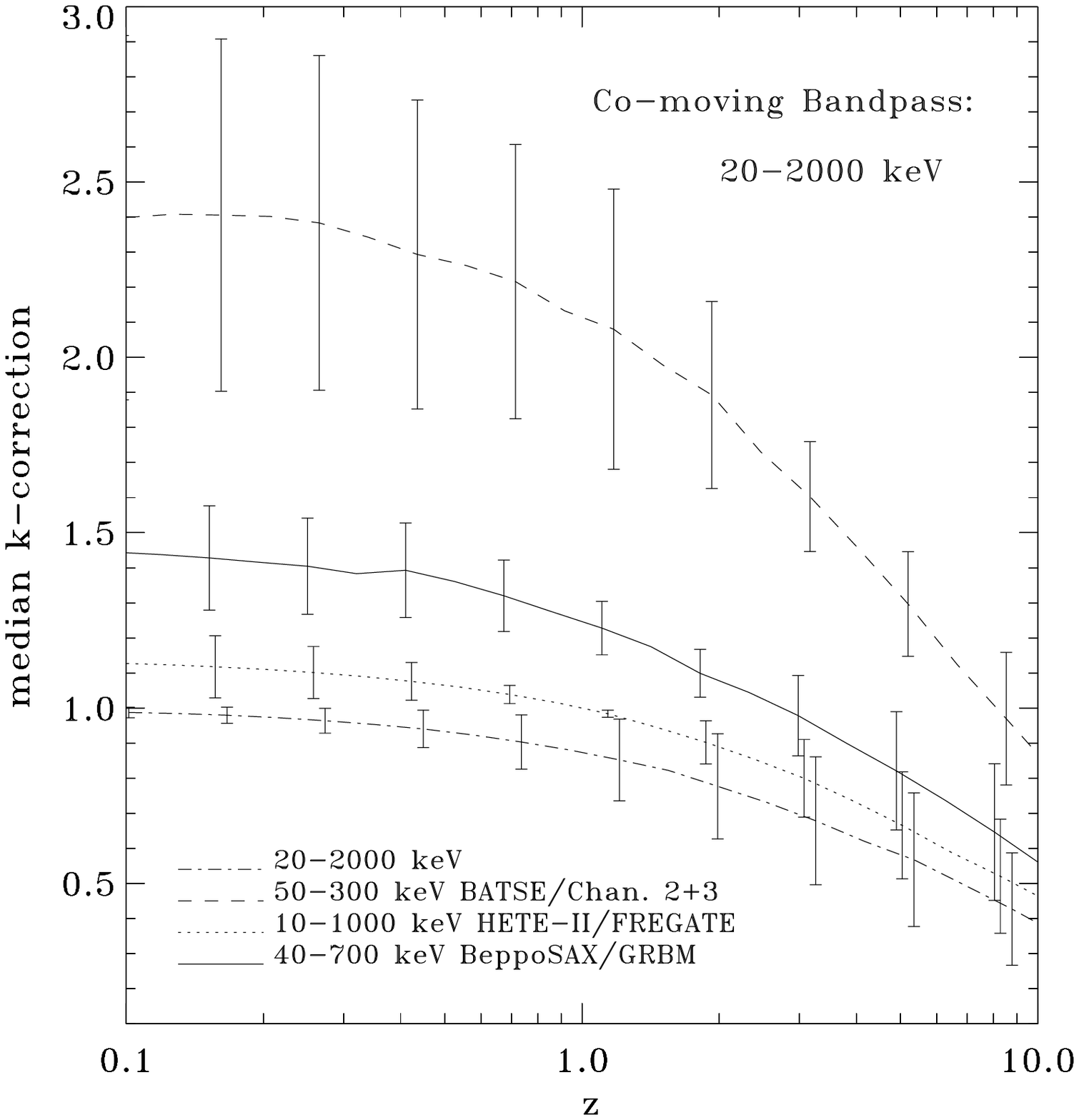,width=3.5in}}
\caption[]{$k$-correction curves for bolometric (left) and co-moving
20--2000 keV bandpasses. The curves shown are the median
$k$-corrections derived using the template spectral method for
observed fluence bandpasses of 20--2000 keV (dash-dotted line),
10--1000 keV (HETE-II/FREGATE; dotted line), 40--700 keV
(BeppoSAX/GRBM; solid line), and 50--300 keV (BATSE/Channels 2+3;
dashed line). In the absence of a reported spectrum, these curves may
be used in conjunction with equation \ref{eq:theken} to estimate the
prompt isotropic-equivalent energy of a burst.  The error bars
represent the 1-$\sigma$ scatter in the ensemble of template spectra,
which are typically grow larger with a smaller observed bandpass
(e.g.~50--300 keV).  Note that the $k$-correction approaches unity and
the scatter shrinks when the observed bandpasses are well-matched to
the redshifted co-moving bandpasses (e.g.~HETE-II/FREGATE near $z \sim
1$ and 20--2000 keV at $z=0$).}  
\label{fig:avgcor}
\end{figure*}

In the future, most GRB fluences will be measured in the energy ranges
40--700 keV (BeppoSAX/GRBM) and 10--1000 keV (HETE-II/FREGATE) and so
it is of interest to know the characteristic $k$-correction for these
instruments.  In figure \ref{fig:avgcor} we show the median
$k$-correction for several observed bandpass energy ranges as a
function of burst redshift and fixed co-moving bandpass.  These median
$k$-correction curves, which are independent of cosmology, were
generated using the template spectral method described herein.  In the
absence of a reported spectrum, these curves may be used
in conjunction with equation \ref{eq:theken} to calculate the prompt
isotropic-equivalent energy of a burst.

\begin{figure*}
\centerline{\psfig{file=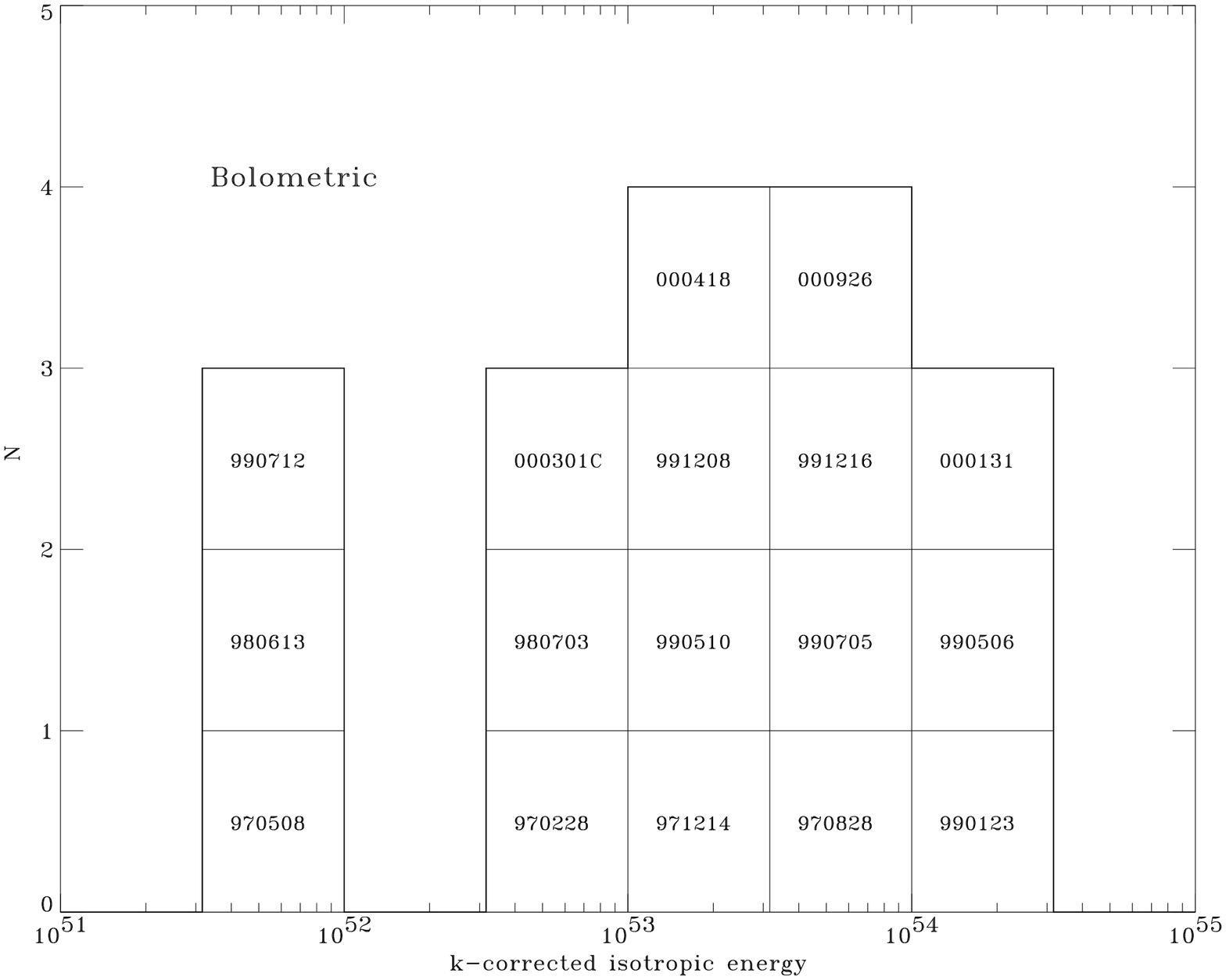,width=4.8in,angle=90}}
\centerline{\psfig{file=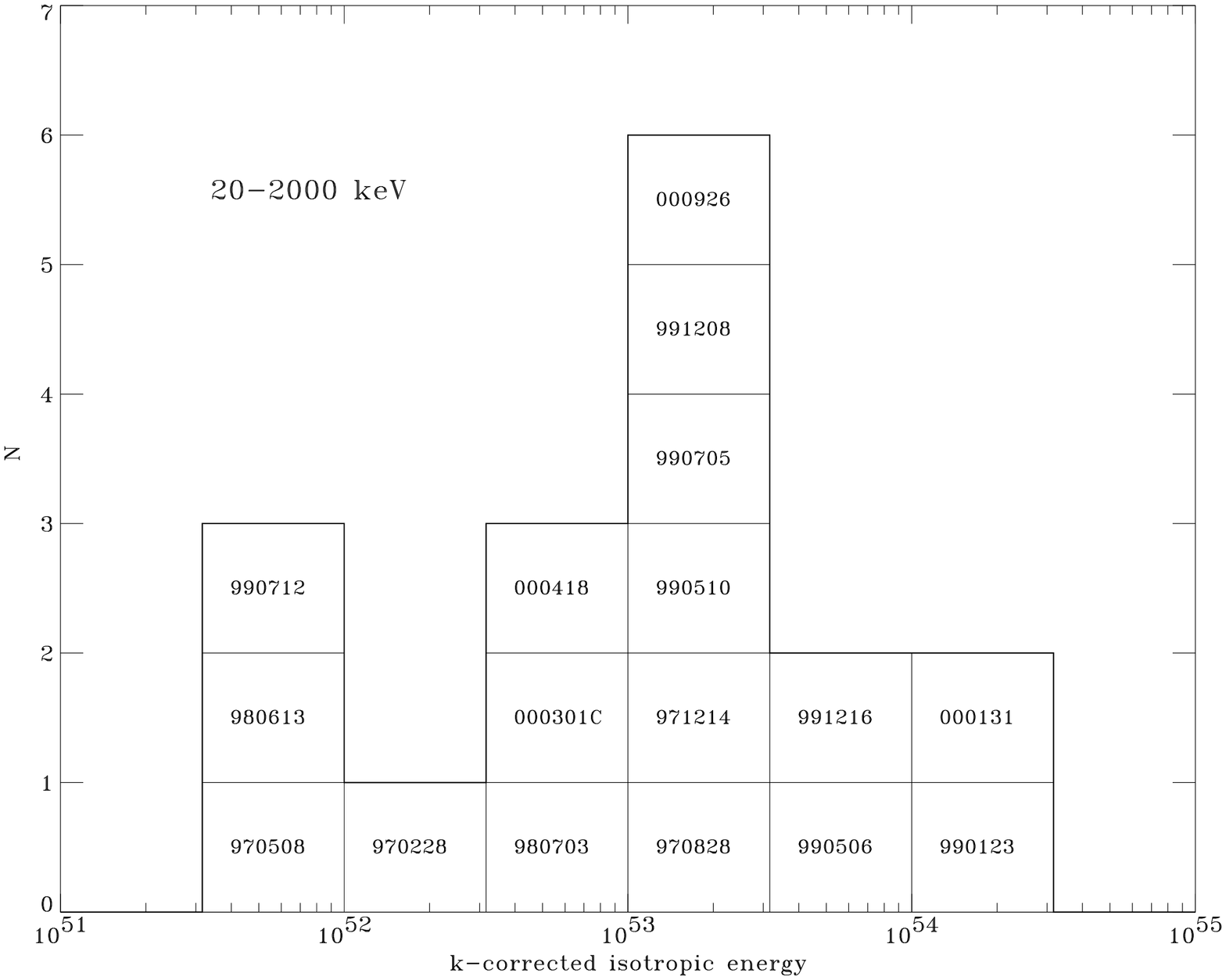,width=4.8in,angle=90}}
\caption[]{Histogram of the bolometric (left) and co-moving 20--2000
keV (right) $k$-corrected prompt energy releases binned with equal log
spacing. The cosmological world model used is $H_0 = 65$ km s$^{-1}$
Mpc$^{-1}$, $\Omega_m = 0.3$, $\Omega_\Lambda$ = 0.7.}
\label{fig:hist-e}
\end{figure*}

Figure \ref{fig:hist-e} shows a histogram of the $k$-corrected
energies of 17 GRBs.  For a cosmology with $H_0 = 65$ km s$^{-1}$
Mpc$^{-1}$, $\Omega_m = 0.3$, $\Omega_\Lambda$ = 0.7, the minimum
bolometric GRB energy of those GRBs with measured redshifts is
\hbox{$6.65 \times 10^{51}$ erg} (GRB 990712) and the maximum
bolometric energy is \hbox{$2.32 \times 10^{54}$} erg (GRB
990123). The median bolometric energy release is \hbox{$2.19 \times
10^{53}$ erg} (GRB 990510) with an r.m.s.~scatter of 0.80 dex.  We
emphasize that this analysis of the characteristic $k$-corrected
energies applies only to the {\it observed} distribution of GRBs with
redshifts; several observational biases (for example, in burst
detection and redshift determination) obscure the true underlying
energy distribution.
 
Given that many GRBs, from analysis of afterglow, are now believed to
be jetted \citepeg{hbf+99,bsf+00,hum+00}, the real energy release of a
given GRB may be substantially less than the isotropic-equivalent
energies derived herein.  To find the real total prompt energy release
one requires an additional correction factor that takes in to account
the geometry of the explosion.  Including geometric corrections, the
distribution of GRB energies appears to tighten to a significantly
narrower distribution than without the geometric corrections (such
work will be presented elsewhere in \citet{fra01}. This tight
distribution, which suggests that GRBs may be viable standard candles,
underscores the ever-increasing need for accurate spectral and fluence
measures as well as an accurate accounting of the order-of-unity
$k$-corrections.

\acknowledgments

The authors thank D.~Band for providing unpublished spectral and
fluence measurements of two recent bursts. We acknowledge helpful
discussions with the members of the Caltech-NRAO-CARA GRB
Collaboration, especially with P.~Price, S.~Kulkarni, and D.~Reichart.
We thank the anonymous referee for helpful comments. JSB gratefully
acknowledges the fellowship from the Fannie and John Hertz Foundation.
The NRAO is a facility of the National Science Foundation operated
under cooperative agreement by Associated Universities, Inc.


\newpage

\begin{deluxetable}{lcccccclrr}
\rotate
\tabletypesize{\scriptsize}
\tablecaption{$k$-corrected GRB energies 20--2000 keV ($H_0 = 65$ km
s$^{-1}$ Mpc$^{-1}$, $\Omega_m = 0.3$, $\Omega_\Lambda$ = 0.7)\label{tab:allspect1}}
\tablehead{
\colhead{Name} &
\colhead{$z$} &
\colhead{$D_l$} &
\multicolumn{3}{c}{Fluence} &
\multicolumn{2}{c}{Energy} &
\colhead{Refs.} \\ 
\colhead{} &
\colhead{} &
\colhead{} &
\colhead{$e_1$} &
\colhead{$e_2$} &
\colhead{$S_{[e_1,e_2]}$} &
\colhead{$E_{[20,2000]}$}~(true) &
\colhead{$E_{[20,2000]}$}~(estimate) &
\colhead{} &
\colhead{} \\
\colhead{} &
\colhead{} &
\colhead{cm} &
\colhead{keV} &
\colhead{keV} &
\colhead{erg cm$^{-2}$} &
\colhead{erg} &
\colhead{erg} &
\colhead{}
}
\startdata
GRB 970228 & 0.6950 &  1.403e+28 
        &     1.5 &   700.0 &   1.17e-05 &   1.42e+52 $\pm$  2.52e+51&  2.05e+52 $\pm$  3.67e+51 & 1, 2 \\
&&      &    50.0 &   300.0 &   3.10e-06 &   7.10e+51 $\pm$  5.56e+50&  1.07e+52 $\pm$  1.90e+51 & 3 \\
&&      &    40.0 &   700.0 &  1.100e-05 &  \ldots &  2.24e+52 $\pm$  2.50e+51 & 4 \\
GRB 970508 & 0.8350 &  1.757e+28 
        &    20.0 &  2000.0 &   3.17e-06 &   5.46e+51 $\pm$  6.12e+50&  6.33e+51 $\pm$  8.22e+50 & 5, 6 \\
&&      &    20.0 &  1000.0 &   3.10e-06 &   7.15e+51 $\pm$  4.65e+50&  7.32e+51 $\pm$  4.63e+50 & 7 \\
&&      &    40.0 &   700.0 &   1.80e-06 &   5.22e+51 $\pm$  8.75e+50&  5.17e+51 $\pm$  8.94e+50 & 4 \\
GRB 970828 & 0.9579 &  2.082e+28 
        &    25.0 &  1800.0 &   7.45e-05 &   2.09e+53 $\pm$  4.81e+51&  2.00e+53 $\pm$  2.17e+52 & 8, 9 \\
&&      &    20.0 &  2000.0 &   9.60e-05 &   2.20e+53 $\pm$  2.40e+52&  2.49e+53 $\pm$  3.41e+52 & 6 \\
GRB 971214 & 3.4180 &  9.877e+28 
        &    20.0 &  2000.0 &   9.44e-06 &   2.11e+53 $\pm$  2.58e+52&  1.85e+53 $\pm$  5.16e+52 & 10, 6 \\
&&      &     2.0 &    12.0 &  3.400e-07 &  \ldots &  3.45e+53 $\pm$  2.79e+53 & 11 \\
&&      &    40.0 &   700.0 &   8.80e-06 &   3.76e+53 $\pm$  6.90e+52&  2.42e+53 $\pm$  4.10e+52 & 4 \\
GRB 980326 & 1.0000 &  2.195e+28 
        &    20.0 &  2000.0 &   9.22e-07 &   3.42e+51 $\pm$  3.74e+50&  2.59e+51 $\pm$  3.61e+50 & 12, 6 \\
&&      &    25.0 &  1800.0 &  1.400e-06 &  \ldots &  4.06e+51 $\pm$  5.91e+50 & 13 \\
GRB 980329 & 1.5000 &  3.625e+28 
        &    20.0 &  2000.0 &   5.51e-05 &   3.18e+53 $\pm$  3.54e+52&  3.19e+53 $\pm$  5.63e+52 & 6 \\
GRB 980425 & 0.0085 &  1.217e+26 
        &    20.0 &  2000.0 &   3.87e-06 &   7.16e+47 $\pm$  7.16e+46&  7.57e+47 $\pm$  7.14e+46 & 6 \\
&&      &    20.0 &  2000.0 &  4.000e-06 &  \ldots &  7.82e+47 $\pm$  1.11e+47 & 14 \\
GRB 980519 & 1.5000 &  3.625e+28 
        &    20.0 &  2000.0 &   1.03e-05 &   7.16e+52 $\pm$  7.99e+51&  5.96e+52 $\pm$  1.05e+52 & 6 \\
&&      &    25.0 &   300.0 &  6.890e-06 &  \ldots &  8.06e+52 $\pm$  1.19e+52 & 15 \\
&&      &    25.0 &  2000.0 &  2.540e-05 &  \ldots &  1.49e+53 $\pm$  3.29e+52 & 16 \\
GRB 980613 & 1.0960 &  2.459e+28 
        &    20.0 &  2000.0 &  1.710e-06 &  \ldots &  5.67e+51 $\pm$  1.04e+51 & 17, 18 \\
GRB 980703 & 0.9660 &  2.103e+28 
        &    20.0 &  2000.0 &   2.26e-05 &   6.01e+52 $\pm$  6.55e+51&  5.94e+52 $\pm$  8.18e+51 & 19, 6 \\
&&      &    20.0 &  2000.0 &  4.590e-05 &  \ldots &  1.21e+53 $\pm$  1.60e+52 & 20 \\
GRB 981220 & 1.5000 &  3.625e+28 
        &    40.0 &   700.0 &  1.000e-05 &  \ldots &  8.12e+52 $\pm$  1.59e+52 & 21 \\
&&      &     2.0 &    12.0 &  1.260e-06 &  \ldots &  3.97e+53 $\pm$  3.08e+53 & 11 \\
GRB 981226 & 1.5000 &  3.625e+28 
        &    40.0 &   700.0 &  4.000e-07 &  \ldots &  3.25e+51 $\pm$  7.86e+50 & 22 \\
GRB 990123 & 1.6000 &  3.925e+28 
        &    20.0 &  2000.0 &   2.68e-04 &   1.44e+54 $\pm$  1.78e+53&  1.73e+54 $\pm$  3.15e+53 & 23, 6 \\
&&      &    20.0 &  2000.0 &  5.090e-04 &  \ldots &  3.28e+54 $\pm$  5.12e+53 & 24 \\
GRB 990506 & 1.3000 &  3.037e+28 
        &    20.0 &  2000.0 &   1.94e-04 &   8.54e+53 $\pm$  1.00e+53&  8.74e+53 $\pm$  1.44e+53 & 25, 6 \\
&&      &    20.0 &  2000.0 &  2.230e-04 &  \ldots &  1.00e+54 $\pm$  1.37e+53 & 26 \\
GRB 990510 & 1.6190 &  3.982e+28 
        &    20.0 &  2000.0 &   2.26e-05 &   1.76e+53 $\pm$  2.00e+52&  1.49e+53 $\pm$  2.73e+52 & 27, 6 \\
&&      &    40.0 &   700.0 &  1.900e-05 &  \ldots &  1.74e+53 $\pm$  1.99e+52 & 28 \\
&&      &    20.0 &  2000.0 &  2.560e-05 &  \ldots &  1.68e+53 $\pm$  2.71e+52 & 29 \\
GRB 990705 & 0.8400 &  1.770e+28 
        &    40.0 &   700.0 &  9.300e-05 &  \ldots &  2.70e+53 $\pm$  2.02e+52 & 30, 31 \\
GRB 990712 & 0.4330 &  7.927e+27 
        &    40.0 &   700.0 &  6.500e-06 &  \ldots &  5.27e+51 $\pm$  6.87e+50 & 27, 32 \\
GRB 991208 & 0.7055 &  1.429e+28 
        &    25.0 &  2000.0 &  1.000e-04 &  \ldots &  1.47e+53 $\pm$  1.98e+52 & 33, 34 \\
GRB 991216 & 1.0200 &  2.250e+28 
        &    20.0 &  2000.0 &   1.94e-04 &   5.35e+53 $\pm$  5.94e+52&  5.64e+53 $\pm$  7.93e+52 & 35, 36 \\
&&      &    20.0 &  2000.0 &  2.000e-04 &  \ldots &  5.82e+53 $\pm$  8.17e+52 & 37 \\
GRB 000131 & 4.5000 &  1.369e+29 
        &    20.0 &  2000.0 &   4.18e-05 &   1.16e+54 $\pm$  1.76e+53&  1.15e+54 $\pm$  3.48e+53 & 38, 36 \\
&&      &    25.0 &   100.0 &  1.000e-05 &  \ldots &  1.47e+54 $\pm$  3.39e+53 & 39 \\
GRB 000301C & 2.0335 &  5.269e+28 
        &    25.0 &  1000.0 &  4.100e-06 &  \ldots &  4.64e+52 $\pm$  6.22e+51 & 40, 41 \\
GRB 000418 & 1.1190 &  2.523e+28 
        &    25.0 &   100.0 &  4.700e-06 &  \ldots &  8.29e+52 $\pm$  3.21e+52 & 42, 43 \\
GRB 000926 & 2.0369 &  5.280e+28 
        &    25.0 &   100.0 &  6.200e-06 &  \ldots &  2.97e+53 $\pm$  9.90e+52 & 44, 45 
\enddata 

\tablecomments{When two references are given in the ``Refs.'' column,
the first is a reference to the redshift, $z$, and the second is the
reference to that particular spectral/fluence measurement.  If only
one reference is given, it is for the that particular spectral/fluence
measurement only.  Redshifts for GRB 980326, GRB 980329, GRB 980519,
GRB 981220, and GRB 981226 are assumed.  The column labeled
``$E_{[20,2000]}$ (true)'' is the $k$-corrected prompt energy release
in the co-moving bandpass 20--2000 keV determined using the spectral
information and fluence given in the reference. The column labeled
``$E_{[20,2000]}$ (estimate)'' is the $k$-corrected prompt energy
release in the co-moving bandpass 20--2000 keV estimated using the
template spectra method described in the text.  If only the
``estimate'' column is given, then that particular reference for the
fluence did not provide spectral information.  If both columns are
given, then the reference did provide spectral and fluence
measurements and the ``estimate'' column is then the energy determined
using the template spectra method and only the provided fluence
measurement.  As seen here (and depicted in figure \ref{fig:encomp}) the
template spectra method adequately recovers the true $k$-corrected
energy.}

\tablerefs{1. \citet{bdk00}; 2. \citet{fac+98}; 3. \citet{pcg+98};
4. \citet{fac+00}; 5. \citet{bdkf98}; 6. \citet{pjb00};
7. \citet{kbp+97}; 8. \citet{dfk+00}; 9. \citet{pbm+00};
10. \citet{kdr+98}; 11. \citet{slb+99}; 12. \citet{bkd+99};
13. \citet{ggv+98c}; 14. \citet{kip+98}; 15. \citet{ihvf99};
16. \citet{con+98}; 17. \citet{dbk00}; 18. \citet{wkc98};
19. \citet{dkb+98b}; 20. \citet{k+98}; 21. \citet{fafc98};
22. \citet{faa+00}; 23. \citet{kdo+99}; 24. \citet{kip+99a};
25. \citet{blo01}; 26. \citet{kip+99b}; 27. \citet{vfk+00};
28. \citet{afc+99}; 29. \citet{kip+99}; 30. Tanvir, priv.~comm.;
31. \citet{afv+00}; 32. \citet{fro00}; 33. \citet{ddb+99};
34. \citet{hcm+00}; 35. \citet{vrh+99}; 36. Band, priv.~comm.;
37. \citet{kip+99c}; 38. \citet{ahp+00}; 39. \citet{hc+00};
40. \citet{cdd+00}; 41. \citet{bsf+00}; 42. \citet{bdd+00};
43. \citet{die01}; 44. \citet{cdk+00}; 45. \citet{phg+00} }

\end{deluxetable}
\begin{deluxetable}{lcccccclrr}
\rotate
\tabletypesize{\scriptsize}
\tablecaption{$k$-corrected GRB bolometric energies ($H_0 = 65$ km
s$^{-1}$ Mpc$^{-1}$, $\Omega_m = 0.3$, $\Omega_\Lambda$ = 0.7)\label{tab:allspect2}}
\tablehead{
\colhead{Name} &
\colhead{$z$} &
\colhead{$D_l$} &
\multicolumn{3}{c}{Fluence} &
\multicolumn{2}{c}{Energy} &
\colhead{Refs.} \\ 
\colhead{} &
\colhead{} &
\colhead{} &
\colhead{$e_1$} &
\colhead{$e_2$} &
\colhead{$S_{[e_1,e_2]}$} &
\colhead{$E_{[0.1,10000]}$}~(true) &
\colhead{$E_{[0.1,10000]}$}~(estimate) &
\colhead{} &
\colhead{} \\
\colhead{} &
\colhead{} &
\colhead{cm} &
\colhead{keV} &
\colhead{keV} &
\colhead{erg cm$^{-2}$} &
\colhead{erg} &
\colhead{erg} &
\colhead{}
}
\startdata
GRB 970228 & 0.6950 &  1.403e+28 
        &     1.5 &   700.0 &   1.17e-05 &   2.68e+52 $\pm$  9.77e+51&  2.56e+52 $\pm$  7.97e+51 & 1, 2 \\
&&      &    50.0 &   300.0 &   3.10e-06 &   7.84e+51 $\pm$  2.56e+51&  1.42e+52 $\pm$  4.64e+51 & 3 \\
&&      &    40.0 &   700.0 &   1.10e-05 &   6.16e+52 $\pm$  1.49e+52&  2.86e+52 $\pm$  8.29e+51 & 4 \\
GRB 970508 & 0.8350 &  1.757e+28 
        &    20.0 &  2000.0 &   3.17e-06 &   9.85e+51 $\pm$  1.94e+51&  8.11e+51 $\pm$  1.28e+51 & 5, 6 \\
&&      &    20.0 &  1000.0 &   3.10e-06 &   9.29e+51 $\pm$  9.09e+50&  9.29e+51 $\pm$  2.01e+51 & 7 \\
&&      &    40.0 &   700.0 &   1.80e-06 &   7.05e+51 $\pm$  1.26e+51&  6.71e+51 $\pm$  2.12e+51 & 4 \\
GRB 970828 & 0.9579 &  2.082e+28 
        &    25.0 &  1800.0 &   7.45e-05 &   2.15e+53 $\pm$  3.22e+52&  2.57e+53 $\pm$  3.33e+52 & 8, 9 \\
&&      &    20.0 &  2000.0 &   9.60e-05 &   3.41e+53 $\pm$  7.90e+52&  3.20e+53 $\pm$  4.88e+52 & 6 \\
GRB 971214 & 3.4180 &  9.877e+28 
        &    20.0 &  2000.0 &   9.44e-06 &   2.81e+53 $\pm$  5.68e+52&  2.78e+53 $\pm$  3.01e+52 & 10, 6 \\
&&      &     2.0 &    12.0 &  3.400e-07 &  \ldots &  5.41e+53 $\pm$  4.91e+53 & 11 \\
&&      &    40.0 &   700.0 &   8.80e-06 &   9.11e+53 $\pm$  4.95e+53&  3.80e+53 $\pm$  6.59e+52 & 12 \\
GRB 980326 & 1.0000 &  2.195e+28 
        &    20.0 &  2000.0 &   9.22e-07 &   4.61e+51 $\pm$  1.03e+51&  3.34e+51 $\pm$  5.03e+50 & 13, 6 \\
&&      &    25.0 &  1800.0 &  1.400e-06 &  \ldots &  5.24e+51 $\pm$  8.43e+50 & 14 \\
GRB 980329 & 1.5000 &  3.625e+28 
        &    20.0 &  2000.0 &   5.51e-05 &   4.19e+53 $\pm$  7.96e+52&  4.21e+53 $\pm$  5.62e+52 & 6 \\
GRB 980425 & 0.0085 &  1.217e+26 
        &    20.0 &  2000.0 &   3.87e-06 &   9.17e+47 $\pm$  1.53e+47&  9.16e+47 $\pm$  1.99e+47 & 6 \\
&&      &    20.0 &  2000.0 &  4.000e-06 &  \ldots &  9.47e+47 $\pm$  2.31e+47 & 15 \\
GRB 980519 & 1.5000 &  3.625e+28 
        &    20.0 &  2000.0 &   1.03e-05 &   8.29e+52 $\pm$  1.24e+52&  7.87e+52 $\pm$  1.05e+52 & 6 \\
&&      &    25.0 &   300.0 &  6.890e-06 &  \ldots &  1.12e+53 $\pm$  3.44e+52 & 16 \\
&&      &    25.0 &  2000.0 &  2.540e-05 &  \ldots &  1.99e+53 $\pm$  3.68e+52 & 17 \\
GRB 980613 & 1.0960 &  2.459e+28 
        &    20.0 &  2000.0 &  1.710e-06 &  \ldots &  7.36e+51 $\pm$  1.34e+51 & 18, 19 \\
GRB 980703 & 0.9660 &  2.103e+28 
        &    20.0 &  2000.0 &   2.26e-05 &   8.09e+52 $\pm$  1.24e+52&  7.66e+52 $\pm$  1.17e+52 & 20, 6 \\
&&      &    20.0 &  2000.0 &  4.590e-05 &  \ldots &  1.56e+53 $\pm$  2.28e+52 & 21 \\
GRB 981220 & 1.5000 &  3.625e+28 
        &    40.0 &   700.0 &  1.000e-05 &  \ldots &  1.09e+53 $\pm$  3.26e+52 & 22 \\
&&      &     2.0 &    12.0 &  1.260e-06 &  \ldots &  4.91e+53 $\pm$  5.97e+53 & 11 \\
GRB 981226 & 1.5000 &  3.625e+28 
        &    40.0 &   700.0 &  4.000e-07 &  \ldots &  4.37e+51 $\pm$  1.46e+51 & 12 \\
GRB 990123 & 1.6000 &  3.925e+28 
        &    20.0 &  2000.0 &   2.68e-04 &   2.32e+54 $\pm$  5.70e+53&  2.31e+54 $\pm$  3.01e+53 & 23, 6 \\
&&      &    20.0 &  2000.0 &  5.090e-04 &  \ldots &  4.38e+54 $\pm$  3.69e+53 & 24 \\
GRB 990506 & 1.3000 &  3.037e+28 
        &    20.0 &  2000.0 &   1.94e-04 &   1.28e+54 $\pm$  1.93e+53&  1.15e+54 $\pm$  1.60e+53 & 25, 6 \\
&&      &    20.0 &  2000.0 &  2.230e-04 &  \ldots &  1.32e+54 $\pm$  1.29e+53 & 26 \\
GRB 990510 & 1.6190 &  3.982e+28 
        &    20.0 &  2000.0 &   2.26e-05 &   2.19e+53 $\pm$  3.45e+52&  1.99e+53 $\pm$  2.59e+52 & 27, 6 \\
&&      &    40.0 &   700.0 &  1.900e-05 &  \ldots &  2.38e+53 $\pm$  5.71e+52 & 28 \\
&&      &    20.0 &  2000.0 &  2.560e-05 &  \ldots &  2.25e+53 $\pm$  2.04e+52 & 29 \\
GRB 990705 & 0.8400 &  1.770e+28 
        &    40.0 &   700.0 &  9.300e-05 &  \ldots &  3.51e+53 $\pm$  9.41e+52 & 30, 31 \\
GRB 990712 & 0.4330 &  7.927e+27 
        &    40.0 &   700.0 &  6.500e-06 &  \ldots &  6.27e+51 $\pm$  1.86e+51 & 27, 32 \\
GRB 991208 & 0.7055 &  1.429e+28 
        &    25.0 &  2000.0 &  1.000e-04 &  \ldots &  1.86e+53 $\pm$  3.07e+52 & 33, 34 \\
GRB 991216 & 1.0200 &  2.250e+28 
        &    20.0 &  2000.0 &   1.94e-04 &   7.83e+53 $\pm$  1.34e+53&  7.29e+53 $\pm$  1.09e+53 & 35, 36 \\
&&      &    20.0 &  2000.0 &  2.000e-04 &  \ldots &  7.52e+53 $\pm$  1.13e+53 & 37 \\
GRB 000131 & 4.5000 &  1.369e+29 
        &    20.0 &  2000.0 &   4.18e-05 &   1.86e+54 $\pm$  4.10e+53&  1.83e+54 $\pm$  2.05e+53 & 38, 36 \\
&&      &    25.0 &   100.0 &  1.000e-05 &  \ldots &  2.42e+54 $\pm$  9.35e+53 & 39 \\
GRB 000301C & 2.0335 &  5.269e+28 
        &    25.0 &  1000.0 &  4.100e-06 &  \ldots &  6.35e+52 $\pm$  1.16e+52 & 40, 41 \\
GRB 000418 & 1.1190 &  2.523e+28 
        &    25.0 &   100.0 &  4.700e-06 &  \ldots &  1.16e+53 $\pm$  4.92e+52 & 42, 43 \\
GRB 000926 & 2.0369 &  5.280e+28 
        &    25.0 &   100.0 &  6.200e-06 &  \ldots &  4.56e+53 $\pm$  1.82e+53 & 44, 45
\enddata 
\tablecomments{See comments of table \ref{tab:allspect1}.}

\tablerefs{1. \citet{bdk00}; 2. \citet{fac+98}; 3. \citet{pcg+98};
4. \citet{fac+00}; 5. \citet{bdkf98}; 6. \citet{pjb00};
7. \citet{kbp+97}; 8. \citet{dfk+00}; 9. \citet{pbm+00};
10. \citet{kdr+98}; 11. \citet{slb+99}; 12. \citet{faa+00};
13. \citet{bkd+99}; 14. \citet{ggv+98c}; 15. \citet{kip+98};
16. \citet{ihvf99}; 17. \citet{con+98}; 18. \citet{dbk00};
19. \citet{wkc98}; 20. \citet{dkb+98b}; 21. \citet{k+98};
22. \citet{fafc98}; 23. \citet{kdo+99}; 24. \citet{kip+99a};
25. \citet{blo01}; 26. \citet{kip+99b}; 27. \citet{vfk+00};
28. \citet{afc+99}; 29. \citet{kip+99}; 30. Tanvir, priv.~comm.;
31. \citet{afv+00}; 32. \citet{fro00}; 33. \citet{ddb+99};
34. \citet{hcm+00}; 35. \citet{vrh+99}; 36. Band, priv.~comm.;
37. \citet{kip+99c}; 38. \citet{ahp+00}; 39. \citet{hc+00};
40. \citet{cdd+00}; 41. \citet{bsf+00}; 42. \citet{bdd+00};
43. \citet{die01}; 44. \citet{cdk+00}; 45. \citet{phg+00}}
\end{deluxetable}

\begin{deluxetable}{lcccccccclrr}
\rotate
\tabletypesize{\scriptsize}
\tablecaption{Summary of $k$-corrected energies in 20--2000 keV for
three world models\label{tab:ensum1}}
\tablehead{
\colhead{Name} &
\colhead{$z$} &
\multicolumn{3}{c}{Fluence} &
\multicolumn{4}{c}{Energy} &
\multicolumn{1}{c}{$k$} &
\colhead{Refs.} \\ 
\colhead{} &
\colhead{} &
\colhead{$S[e_1, e_2]$} &
\colhead{$e_1$} &
\colhead{$e_2$} &
\colhead{$E$} &
\colhead{$E_{[20,2000]}$} &
\colhead{$E_{[20,2000]}$} &
\colhead{$E_{[20,2000]}$} &
\colhead{} &
\colhead{} \\
\colhead{} &
\colhead{} &
\colhead{erg cm$^{-2}$} &
\colhead{keV} &
\colhead{keV} &
\colhead{0.3,0.7} &
\colhead{0.3,0.7} &
\colhead{0.05,0.0} &
\colhead{1.0,0.0} &
\colhead{} &
\colhead{} &
}
\startdata
   970228 &   0.6950 &  1.170e-05 &     1.5 &   700.0 &  1.71e+52 &  1.42e+52 &  1.25e+52 &  9.00e+51 &    0.83 & 1 \\
   970508 &   0.8350 &  3.170e-06 &    20.0 &  2000.0 &  6.70e+51 &  5.46e+51 &  4.88e+51 &  3.30e+51 &    0.81 & 2 \\
   970828 &   0.9579 &  9.600e-05 &    20.0 &  2000.0 &  2.67e+53 &  2.20e+53 &  2.00e+53 &  1.28e+53 &    0.82 & 2 \\
   971214 &   3.4180 &  9.440e-06 &    20.0 &  2000.0 &  2.62e+53 &  2.11e+53 &  3.29e+53 &  9.38e+52 &    0.80 & 2 \\
   980326 &   1.0000 &  9.220e-07 &    20.0 &  2000.0 &  2.79e+51 &  3.42e+51 &  3.13e+51 &  1.97e+51 &    1.23 & 2 \\
   980329 &   1.5000 &  5.510e-05 &    20.0 &  2000.0 &  3.64e+53 &  3.18e+53 &  3.20e+53 &  1.65e+53 &    0.87 & 2 \\
   980425 &   0.0085 &  3.870e-06 &    20.0 &  2000.0 &  7.15e+47 &  7.16e+47 &  7.12e+47 &  7.10e+47 &    1.00 & 2 \\
   980519 &   1.5000 &  1.030e-05 &    20.0 &  2000.0 &  6.80e+52 &  7.16e+52 &  7.22e+52 &  3.73e+52 &    1.05 & 2 \\
   980613 &   1.0960 &  1.710e-06 &    20.0 &  2000.0 &  6.20e+51 &  5.67e+51 &  5.27e+51 &  3.19e+51 &    0.91 & 3 \\
   980703 &   0.9660 &  2.260e-05 &    20.0 &  2000.0 &  6.39e+52 &  6.01e+52 &  5.47e+52 &  3.50e+52 &    0.94 & 2 \\
   981220 &   1.5000 &  1.000e-05 &    40.0 &   700.0 &  6.60e+52 &  8.12e+52 &  8.19e+52 &  4.23e+52 &    1.23 & 4 \\
   981226 &   1.5000 &  4.000e-07 &    40.0 &   700.0 &  2.64e+51 &  3.25e+51 &  3.27e+51 &  1.69e+51 &    1.23 & 5 \\
   990123 &   1.6000 &  2.680e-04 &    20.0 &  2000.0 &  2.00e+54 &  1.44e+54 &  1.48e+54 &  7.37e+53 &    0.72 & 2 \\
   990506 &   1.3000 &  1.940e-04 &    20.0 &  2000.0 &  9.78e+53 &  8.54e+53 &  8.25e+53 &  4.60e+53 &    0.87 & 2 \\
   990510 &   1.6190 &  2.260e-05 &    20.0 &  2000.0 &  1.72e+53 &  1.76e+53 &  1.83e+53 &  9.02e+52 &    1.03 & 2 \\
   990705 &   0.8400 &  9.300e-05 &    40.0 &   700.0 &  1.99e+53 &  2.70e+53 &  2.42e+53 &  1.63e+53 &    1.36 & 6 \\
   990712 &   0.4330 &  6.500e-06 &    40.0 &   700.0 &  3.58e+51 &  5.27e+51 &  4.65e+51 &  3.78e+51 &    1.47 & 7 \\
   991208 &   0.7055 &  1.000e-04 &    25.0 &  2000.0 &  1.50e+53 &  1.47e+53 &  1.30e+53 &  9.29e+52 &    0.98 & 8 \\
   991216 &   1.0200 &  1.940e-04 &    20.0 &  2000.0 &  6.11e+53 &  5.35e+53 &  4.91e+53 &  3.07e+53 &    0.88 & 9 \\
   000131 &   4.5000 &  4.180e-05 &    20.0 &  2000.0 &  1.79e+54 &  1.16e+54 &  2.25e+54 &  4.98e+53 &    0.65 & 9 \\
  000301C &   2.0335 &  4.100e-06 &    25.0 &  1000.0 &  4.71e+52 &  4.64e+52 &  5.28e+52 &  2.26e+52 &    0.98 & 10 \\
   000418 &   1.1190 &  4.700e-06 &    25.0 &   100.0 &  1.77e+52 &  8.29e+52 &  7.74e+52 &  4.64e+52 &    4.67 & 11 \\
   000926 &   2.0369 &  6.200e-06 &    25.0 &   100.0 &  7.15e+52 &  2.97e+53 &  3.38e+53 &  1.44e+53 &    4.15 & 12 
\enddata 

\tablecomments{The column marked $E$ is the ``uncorrected'' energy
found using the equation $E = 4 \pi D_l^2 S_{[e_1,e_1]}/(1+z)$.  The
columns marked with $E_{[20,2000]}$ are properly $k$-corrected
energies using the methodology described in the text. Each energy
column is for a specific world model represented by ($\Omega_m$,
$\Omega_\Lambda$); all world models use $H_0 = 65$ km s$^{-1}$
Mpc$^{-1}$. The ``$k$'' column contains the $k$-corrections which is
the ratio of the $k$-corrected energy to the uncorrected energy for
$\Omega_m=0.3$, $\Omega_\Lambda=0.7$. Note that most $k$-corrections
are of order unity except in a few cases where the reported fluence is
in a narrow bandpass (e.g.~GRB 000926). References are for the
spectral/fluence measurements only; references for the redshift
determination can be found table \ref{tab:allspect1}.}

\tablerefs{ 1. \citet{fac+98}; 2. \citet{pjb00}; 3. \citet{wkc98};
4. \citet{fafc98}; 5. \citet{faa+00}; 6. \citet{afv+00};
7. \citet{fro00}; 8. \citet{hcm+00}; 9. Band, priv.~comm.;
10. \citet{bsf+00}; 11. \citet{die01}; 12. \citet{phg+00}}

\end{deluxetable}
\begin{deluxetable}{lcccccccclrr}
\rotate
\tabletypesize{\scriptsize}
\tablecaption{Summary of $k$-corrected bolometric energies for
three world models\label{tab:ensum2}}
\tablehead{
\colhead{Name} &
\colhead{$z$} &
\multicolumn{3}{c}{Fluence} &
\multicolumn{4}{c}{Energy} &
\multicolumn{1}{c}{$k$} &
\colhead{Refs.} \\ 
\colhead{} &
\colhead{} &
\colhead{$S[e_1, e_2]$} &
\colhead{$e_1$} &
\colhead{$e_2$} &
\colhead{$E$} &
\colhead{$E_{[0.1,10000]}$} &
\colhead{$E_{[0.1,10000]}$} &
\colhead{$E_{[0.1,10000]}$} &
\colhead{} &
\colhead{} \\
\colhead{} &
\colhead{} &
\colhead{erg cm$^{-2}$} &
\colhead{keV} &
\colhead{keV} &
\colhead{0.3,0.7} &
\colhead{0.3,0.7} &
\colhead{0.05,0.0} &
\colhead{1.0,0.0} &
\colhead{} &
\colhead{} &
}
\startdata
    970228 &   0.6950 &  1.170e-05 &     1.5 &   700.0 &  1.71e+52 &  2.68e+52 &  2.36e+52 &  1.70e+52 &    1.57 & 1 \\
   970508 &   0.8350 &  3.170e-06 &    20.0 &  2000.0 &  6.70e+51 &  9.85e+51 &  8.81e+51 &  5.96e+51 &    1.47 & 2 \\
   970828 &   0.9579 &  9.600e-05 &    20.0 &  2000.0 &  2.67e+53 &  3.41e+53 &  3.10e+53 &  1.99e+53 &    1.28 & 2 \\
   971214 &   3.4180 &  9.440e-06 &    20.0 &  2000.0 &  2.62e+53 &  2.81e+53 &  4.38e+53 &  1.25e+53 &    1.07 & 2 \\
   980326 &   1.0000 &  9.220e-07 &    20.0 &  2000.0 &  2.79e+51 &  4.61e+51 &  4.22e+51 &  2.66e+51 &    1.65 & 2 \\
   980329 &   1.5000 &  5.510e-05 &    20.0 &  2000.0 &  3.64e+53 &  4.19e+53 &  4.23e+53 &  2.18e+53 &    1.15 & 2 \\
   980425 &   0.0085 &  3.870e-06 &    20.0 &  2000.0 &  7.15e+47 &  9.17e+47 &  9.13e+47 &  9.09e+47 &    1.28 & 2 \\
   980519 &   1.5000 &  1.030e-05 &    20.0 &  2000.0 &  6.80e+52 &  8.29e+52 &  8.36e+52 &  4.31e+52 &    1.22 & 2 \\
   980613 &   1.0960 &  1.710e-06 &    20.0 &  2000.0 &  6.20e+51 &  7.80e+51 &  7.25e+51 &  4.39e+51 &    1.26 & 3 \\
   980703 &   0.9660 &  2.260e-05 &    20.0 &  2000.0 &  6.39e+52 &  8.09e+52 &  7.36e+52 &  4.71e+52 &    1.27 & 2 \\
   981220 &   1.5000 &  1.000e-05 &    40.0 &   700.0 &  6.60e+52 &  1.16e+53 &  1.17e+53 &  6.02e+52 &    1.75 & 4 \\
   981226 &   1.5000 &  4.000e-07 &    40.0 &   700.0 &  2.64e+51 &  4.63e+51 &  4.67e+51 &  2.41e+51 &    1.75 & 5 \\
   990123 &   1.6000 &  2.680e-04 &    20.0 &  2000.0 &  2.00e+54 &  2.32e+54 &  2.39e+54 &  1.19e+54 &    1.16 & 2 \\
   990506 &   1.3000 &  1.940e-04 &    20.0 &  2000.0 &  9.78e+53 &  1.28e+54 &  1.24e+54 &  6.90e+53 &    1.31 & 2 \\
   990510 &   1.6190 &  2.260e-05 &    20.0 &  2000.0 &  1.72e+53 &  2.19e+53 &  2.27e+53 &  1.12e+53 &    1.27 & 2 \\
   990705 &   0.8400 &  9.300e-05 &    40.0 &   700.0 &  1.99e+53 &  3.72e+53 &  3.33e+53 &  2.25e+53 &    1.87 & 6 \\
   990712 &   0.4330 &  6.500e-06 &    40.0 &   700.0 &  3.58e+51 &  6.65e+51 &  5.87e+51 &  4.77e+51 &    1.86 & 7 \\
   991208 &   0.7055 &  1.000e-04 &    25.0 &  2000.0 &  1.50e+53 &  1.97e+53 &  1.74e+53 &  1.25e+53 &    1.31 & 8 \\
   991216 &   1.0200 &  1.940e-04 &    20.0 &  2000.0 &  6.11e+53 &  7.83e+53 &  7.19e+53 &  4.49e+53 &    1.28 & 9 \\
   000131 &   4.5000 &  4.180e-05 &    20.0 &  2000.0 &  1.79e+54 &  1.86e+54 &  3.60e+54 &  7.98e+53 &    1.04 & 9 \\
  000301C &   2.0335 &  4.100e-06 &    25.0 &  1000.0 &  4.71e+52 &  6.73e+52 &  7.67e+52 &  3.28e+52 &    1.43 & 10 \\
   000418 &   1.1190 &  4.700e-06 &    25.0 &   100.0 &  1.77e+52 &  1.23e+53 &  1.15e+53 &  6.90e+52 &    6.94 & 11 \\
   000926 &   2.0369 &  6.200e-06 &    25.0 &   100.0 &  7.15e+52 &  4.83e+53 &  5.51e+53 &  2.35e+53 &    6.75 & 12
\enddata
\tablecomments{See notes from table \ref{tab:ensum1}.}
\tablerefs{1. \citet{fac+98}; 
2. \citet{pjb00}; 
3. \citet{wkc98}; 
4. \citet{fafc98}; 
5. \citet{faa+00}; 
6. \citet{afv+00}; 
7. \citet{fro00}; 
8. \citet{hcm+00}; 
9. Band, priv.~comm.; 
10. \citet{bsf+00}; 
11. \citet{die01}; 
12. \citet{phg+00}}

\end{deluxetable}


\begin{thebibliography}{60}
\expandafter\ifx\csname natexlab\endcsname\relax\def\natexlab#1{#1}\fi

\bibitem[{{Amati} {et~al.}(2000){Amati}, {Frontera}, {Vietri}, {in't Zand},
  {Soffitta}, {Costa}, {del Sordo}, {Pian}, {Piro}, {Antonelli}, {dal Fiume},
  {Feroci}, {Gandolfi}, {Guidorzi}, {Heise}, {Kuulkers}, {Masetti},
  {Montanari}, {Nicastro}, {Orlandini}, \& {Palazzi}}]{afv+00}
{Amati}, L. {\it et al.}, 2000, Science, 290, 953

\bibitem[{Amati {et~al.}(1999)}]{afc+99}
Amati, L. {\it et al.}, 1999, {GCN} notice 317

\bibitem[{{Andersen} {et~al.}(2000){Andersen}, {Hjorth}, {Pedersen}, {Jensen},
  {Hunt}, {Gorosabel}, {M{\o}ller}, {Fynbo}, {Kippen}, {Thomsen}, {Olsen},
  {Christensen}, {Vestergaard}, {Masetti}, {Palazzi}, {Hurley}, {Cline},
  {Kaper}, \& {Jaunsen}}]{ahp+00}
{Andersen}, M.~I. {\it et al.}, 2000, A\&A, 364, L54

\bibitem[{{Band} {et~al.}(1993){Band}, {Matteson}, {Ford}, {Schaefer},
  {Palmer}, {Teegarden}, {Cline}, {Briggs}, {Paciesas}, {Pendleton}, {Fishman},
  {Kouveliotou}, {Meegan}, {Wilson}, \& {Lestrade}}]{bmf+93}
{Band}, D. {\it et al.}, 1993, ApJ, 413, 281

\bibitem[{{Berger} {et~al.}(2000){Berger}, {Sari}, {Frail}, {Kulkarni},
  {Bertoldi}, {Peck}, {Menten}, {Shepherd}, {Moriarty-Schieven}, {Pooley},
  {Bloom}, {Diercks}, {Galama}, \& {Hurley}}]{bsf+00}
{Berger}, E. {\it et al.}, 2000, Apj, 545, 56

\bibitem[{Berger {et~al.}(2001)}]{die01}
Berger, E. {\it et al.}, 2001, {ApJ}, submitted; astro-ph/0102282

\bibitem[{Bloom {et~al.}(2000{\natexlab{a}})Bloom, Djorgovski, \&
  Kulkarni}]{bdk00}
Bloom, J.~S., Djorgovski, S.~G., and Kulkarni, S.~R. 2000{\natexlab{a}},
  submitted to ApJ. astro-ph/0007244

\bibitem[{Bloom {et~al.}(1998)Bloom, Djorgovski, Kulkarni, \& Frail}]{bdkf98}
Bloom, J.~S., Djorgovski, S.~G., Kulkarni, S.~R., and Frail, D.~A. 1998, ApJ,
  507, L25

\bibitem[{{Bloom} {et~al.}(1996){Bloom}, {Fenimore}, \& {in 't Zand}}]{bfi96}
{Bloom}, J.~S., {Fenimore}, E.~E., and {in 't Zand}, J. 1996, in 3rd Hunstville
  Symposium of Gamma-Ray Bursts, 321

\bibitem[{Bloom {et~al.}(1999)}]{bkd+99}
Bloom, J.~S. {\it et al.}, 1999, Nature, 401, 453

\bibitem[{Bloom {et~al.}(2000{\natexlab{b}})}]{bdd+00}
---. 2000{\natexlab{b}}, {GCN} notice 661

\bibitem[{Bloom {et~al.}(2000{\natexlab{c}})}]{blo01}
---. 2000{\natexlab{c}}, in preparation

\bibitem[{Castro {et~al.}(2000{\natexlab{a}})Castro, Diercks, Djorgovski,
  Kulkarni, Galama, Bloom, Harrison, \& Frail}]{cdd+00}
Castro, S.~M., Diercks, A., Djorgovski, S.~G., Kulkarni, S.~R., Galama, T.~J.,
  Bloom, J.~S., Harrison, F.~A., and Frail, D.~A. 2000{\natexlab{a}}, {GCN}
  notice 605

\bibitem[{Castro {et~al.}(2000{\natexlab{b}})Castro, Djorgovski, Kulkarni,
  Bloom, Galama, Harrison, \& Frail}]{cdk+00}
Castro, S.~M., Djorgovski, S.~G., Kulkarni, S.~R., Bloom, J.~S., Galama, T.~J.,
  Harrison, F.~A., and Frail, D.~A. 2000{\natexlab{b}}, {GCN} notice 851

\bibitem[{Connaughton(1998)}]{con+98}
Connaughton, V. 1998, {GCN} notice 86

\bibitem[{Djorgovski {et~al.}(2000{\natexlab{a}})Djorgovski, Bloom, \&
  Kulkarni}]{dbk00}
Djorgovski, S.~G., Bloom, J.~S., and Kulkarni, S.~R. 2000{\natexlab{a}}, {ApJ
  Lett.}, accepted; astro-ph/0008029

\bibitem[{Djorgovski {et~al.}(1999)Djorgovski, Dierks, Bloom, Kulkarni,
  Filippenko, Hillenbrand, \& Carpenter}]{ddb+99}
Djorgovski, S.~G., Dierks, A., Bloom, J.~S., Kulkarni, S.~R., Filippenko,
  A.~V., Hillenbrand, L.~A., and Carpenter, J. 1999, {GCN} notice 481

\bibitem[{Djorgovski {et~al.}(2000{\natexlab{b}})Djorgovski, Frail, Kulkarni,
  Bloom, Odewahn, \& Diercks}]{dfk+00}
Djorgovski, S.~G., Frail, D.~A., Kulkarni, S.~R., Bloom, J., Odewahn, S.~C.,
  and Diercks, A. 2000{\natexlab{b}}, {ApJ} (Let) submitted

\bibitem[{{Djorgovski} {et~al.}(1998){Djorgovski}, {Kulkarni}, {Bloom},
  {Goodrich}, {Frail}, {Piro}, \& {Palazzi}}]{dkb+98b}
{Djorgovski}, S.~G., {Kulkarni}, S.~R., {Bloom}, J.~S., {Goodrich}, R.,
  {Frail}, D.~A., {Piro}, L., and {Palazzi}, E. 1998, ApJ, 508, L17

\bibitem[{{Fenimore} \& {Bloom}(1995)}]{fb95}
{Fenimore}, E.~E. and {Bloom}, J.~S. 1995, ApJ, 453, 25

\bibitem[{{Fenimore} {et~al.}(1993){Fenimore}, {Epstein}, {Ho}, {Klebesadel},
  {Lacey}, {Laros}, {Meier}, {Strohmayer}, {Pendleton}, {Fishman},
  {Kouveliotou}, \& {Meegan}}]{feh+93}
{Fenimore}, E.~E. {\it et al.}, 1993, Nature, 366, 40

\bibitem[{Frail {et~al.}(2001)}]{fra01}
Frail, D.~A. {\it et al.}, 2001, submitted to Nature; astro-ph/0102282

\bibitem[{{Frontera} {et~al.}(1998){Frontera}, {Amati}, {Costa}, {Feroci},
  {Muller}, {Pizzichini}, {Cinti}, {dal Fiume}, {Heise}, {Nicastro},
  {Orlandini}, {Palazzi}, \& {in 't Zand}}]{fac+98}
{Frontera}, F. {\it et al.}, 1998, in Gamma-Ray Bursts, 446

\bibitem[{{Frontera} {et~al.}(2000{\natexlab{a}}){Frontera}, {Amati}, {Costa},
  {Muller}, {Pian}, {Piro}, {Soffitta}, {Tavani}, {Castro-Tirado}, {Dal Fiume},
  {Feroci}, {Heise}, {Masetti}, {Nicastro}, {Orlandini}, {Palazzi}, \&
  {Sari}}]{fac+00}
{Frontera}, F. {\it et al.}, 2000{\natexlab{a}}, ApJS, 127, 59

\bibitem[{Frontera {et~al.}(1998)Frontera, Amati, Feroci, \& Costa}]{fafc98}
Frontera, F., Amati, L., Feroci, M., and Costa, E. 1998, {GCN} notice 167

\bibitem[{{Frontera} {et~al.}(2000{\natexlab{b}}){Frontera}, {Antonelli},
  {Amati}, {Montanari}, {Costa}, {Dal Fiume}, {Giommi}, {Feroci}, {Gennaro},
  {Heise}, {Masetti}, {Muller}, {Nicastro}, {Orlandini}, {Palazzi}, {Pian},
  {Piro}, {Soffitta}, {Stornelli}, {in 't Zand}, {Frail}, {Kulkarni}, \&
  {Vietri}}]{faa+00}
{Frontera}, F. {\it et al.}, 2000{\natexlab{b}}, ApJ, 540, 697

\bibitem[{Frontera {et~al.}(2000)}]{fro00}
Frontera, F. {\it et al.}, 2000, {ApJ} (Let), submitted

\bibitem[{{Groot} {et~al.}(1998){Groot}, {Galama}, {Vreeswijk}, {Wijers},
  {Pian}, {Palazzi}, {Van Paradijs}, {Kouveliotou}, {In 't Zand}, {Heise},
  {Robinson}, {Tanvir}, {Lidman}, {Tinney}, {Keane}, {Briggs}, {Hurley},
  {Gonzalez}, {Hall}, {Smith}, {Covarrubias}, {Jonker}, {Casares}, {Frontera},
  {Feroci}, {Piro}, {Costa}, {Smith}, {Jones}, {Windridge}, {Bland-Hawthorn},
  {Veilleux}, {Garcia}, {Brown}, {Stanek}, {Castro-Tirado}, {Gorosabel},
  {Greiner}, {Jaeger}, {Bohm}, \& {Fricke}}]{ggv+98c}
{Groot}, P.~J. {\it et al.}, 1998, ApJ, 502, L123

\bibitem[{{Hakkila} {et~al.}(1996){Hakkila}, {Meegan}, {Horack}, {Pendleton},
  {Briggs}, {Mallozzi}, {Koshut}, {Preece}, \& {Paciesas}}]{hmh+96}
{Hakkila}, J. {\it et al.}, 1996, ApJ, 462, 125

\bibitem[{{Halpern} {et~al.}(2000){Halpern}, {Uglesich}, {Mirabal}, {Kassin},
  {Thorstensen}, {Keel}, {Diercks}, {Bloom}, {Harrison}, {Mattox}, \&
  {Eracleous}}]{hum+00}
{Halpern}, J.~P. {\it et al.}, 2000, ApJ, 543, 697

\bibitem[{{Harrison} {et~al.}(1999){Harrison}, {Bloom}, {Frail}, {Sari},
  {Kulkarni}, {Djorgovski}, {Axelrod}, {Mould}, {Schmidt}, {Wieringa}, {Wark},
  {Subrahmanyan}, {McConnell}, {McCarthy}, {Schaefer}, {McMahon}, {Markze},
  {Firth}, {Soffitta}, \& {Amati}}]{hbf+99}
{Harrison}, F.~A. {\it et al.}, 1999, ApJ, 523, L121

\bibitem[{{Hurley} {et~al.}(2000){Hurley}, {Cline}, {Mazets}, {Aptekar},
  {Golenetskii}, {Frederiks}, {Frail}, {Kulkarni}, {Trombka}, {McClanahan},
  {Starr}, \& {Goldsten}}]{hcm+00}
{Hurley}, K. {\it et al.}, 2000, ApJ, 534, L23

\bibitem[{Hurley {et~al.}(2000)}]{hc+00}
Hurley, K. {\it et al.}, 2000, {GCN} notice 529

\bibitem[{{in 't Zand} {et~al.}(1999){in 't Zand}, {Heise}, {van Paradijs}, \&
  {Fenimore}}]{ihvf99}
{in 't Zand}, J. J.~M., {Heise}, J., {van Paradijs}, J., and {Fenimore}, E.~E.
  1999, ApJ, 516, L57

\bibitem[{Kippen {et~al.}(1998{\natexlab{a}})}]{kip+98}
Kippen, R.~M. {\it et al.}, 1998{\natexlab{a}}, {GCN} notice 67

\bibitem[{Kippen {et~al.}(1998{\natexlab{b}})}]{k+98}
---. 1998{\natexlab{b}}, {GCN} notice 143

\bibitem[{Kippen {et~al.}(1999{\natexlab{a}})}]{kip+99a}
---. 1999{\natexlab{a}}, {GCN} 224

\bibitem[{Kippen {et~al.}(1999{\natexlab{b}})}]{kip+99b}
---. 1999{\natexlab{b}}, {GCN} 306

\bibitem[{Kippen {et~al.}(1999{\natexlab{c}})}]{kip+99}
---. 1999{\natexlab{c}}, {GCN} 322

\bibitem[{Kippen {et~al.}(1999{\natexlab{d}})}]{kip+99c}
---. 1999{\natexlab{d}}, {GCN} 463

\bibitem[{{Kommers} {et~al.}(2000){Kommers}, {Lewin}, {Kouveliotou}, {van
  Paradijs}, {Pendleton}, {Meegan}, \& {Fishman}}]{klk+00}
{Kommers}, J.~M., {Lewin}, W. H.~G., {Kouveliotou}, C., {van Paradijs}, J.,
  {Pendleton}, G.~N., {Meegan}, C.~A., and {Fishman}, G.~J. 2000, ApJ, 533, 696

\bibitem[{Kouveliotou {et~al.}(1997)Kouveliotou, Briggs, Preece, Fishman,
  Meegan, \& Harmon}]{kbp+97}
Kouveliotou, C., Briggs, M.~S., Preece, R., Fishman, G.~J., Meegan, C.~A., and
  Harmon, B.~A. 1997.
\newblock IAU circular 6660

\bibitem[{{Kulkarni} {et~al.}(1998){Kulkarni}, {Djorgoski}, {Ramaprakash},
  {Goodrich}, {Bloom}, {Adelberger}, {Kundic}, {Lubin}, {Frail}, {Frontera},
  {Feroci}, {Nicastro}, {Barth}, {Davis}, {Filippenko}, \& {Newman}}]{kdr+98}
{Kulkarni}, S.~R. {\it et al.}, 1998, Nature, 393, 35

\bibitem[{{Kulkarni} {et~al.}(1999){Kulkarni}, {Djorgovski}, {Odewahn},
  {Bloom}, {Gal}, {Koresko}, {Harrison}, {Lubin}, {Armus}, {Sari},
  {Illingworth}, {Kelson}, {Magee}, {Van Dokkum}, {Frail}, {Mulchaey},
  {Malkan}, {MCClean}, {Teplitz}, {Koerner}, {Kirkpatrick}, {Kobayashi},
  {Yadigaroglu}, {Halpern}, {Piran}, {Goodrich}, {Chaffee}, {Feroci}, \&
  {Costa}}]{kdo+99}
---. 1999, Nature, 398, 389

\bibitem[{Mallozzi {et~al.}(1996)Mallozzi, Pendleton, \& Paciesas}]{mpp96}
Mallozzi, R.~S., Pendleton, G.~N., and Paciesas, W.~S. 1996, ApJ, 471, 636

\bibitem[{Meegan {et~al.}(1992)}]{mfw+92}
Meegan, C. {\it et al.}, 1992, Nature, 355, 143

\bibitem[{Metzger {et~al.}(1997)Metzger, Djorgovski, Kulkarni, Steidel,
  Adelberger, Frail, Costa, \& Fronterra}]{mdk+97}
Metzger, M.~R., Djorgovski, S.~G., Kulkarni, S.~R., Steidel, C.~C., Adelberger,
  K.~L., Frail, D.~A., Costa, E., and Fronterra, F. 1997, Nature, 387, 879

\bibitem[{{Palmer} {et~al.}(1998){Palmer}, {Cline}, {Gehrels}, {Hurley},
  {Kurczynski}, {Madden}, {Pehl}, {Ramaty}, {Seifert}, \& {Teegarden}}]{pcg+98}
{Palmer}, D.~M. {\it et al.}, 1998, in Gamma-Ray Bursts, 304

\bibitem[{{Piran}(1992)}]{pir92}
{Piran}, T. 1992, ApJ, 389, L45

\bibitem[{Piran {et~al.}(2000)Piran, Jimenez, \& Band}]{pjb00}
Piran, T., Jimenez, R., and Band, D. 2000, in Gamma Ray Bursts: 5th Huntsville
  Symposium (Woodbury, New York: AIP), {87--91}

\bibitem[{{Preece} {et~al.}(2000){Preece}, {Briggs}, {Mallozzi}, {Pendleton},
  {Paciesas}, \& {Band}}]{pbm+00}
{Preece}, R.~D., {Briggs}, M.~S., {Mallozzi}, R.~S., {Pendleton}, G.~N.,
  {Paciesas}, W.~S., and {Band}, D.~L. 2000, ApJS, 126, 19

\bibitem[{{Preece} {et~al.}(1998){Preece}, {Pendleton}, {Briggs}, {Mallozzi},
  {Paciesas}, {Band}, {Matteson}, \& {Meegan}}]{ppb+98}
{Preece}, R.~D., {Pendleton}, G.~N., {Briggs}, M.~S., {Mallozzi}, R.~S.,
  {Paciesas}, W.~S., {Band}, D.~L., {Matteson}, J.~L., and {Meegan}, C.~A.
  1998, ApJ, 496, 849

\bibitem[{Price {et~al.}(2000)Price, Harrison, Galama, Reichart, Axelrod,
  Berger, Bloom, Busche, Cline, Diercks, Djorgovski, Frail, Gal-Yam, Halpern,
  Holtzman, Hunt, Hurley, Jacoby, Kimble, Kulkarni, Mirabal, Morrison, Ofek,
  Pevunova, Sari, Schmidt, Turnshek, \& Yost}]{phg+00}
Price, P.~A. {\it et al.}, 2000, {ApJ} (Let) accepted, astro-ph/0012303

\bibitem[{{Schmidt}(1999)}]{sch99}
{Schmidt}, M. 1999, \apjl, 523, L117

\bibitem[{{Smith} {et~al.}(1999){Smith}, {Levine}, {Bradt}, {Remillard},
  {Jernigan}, {Hurley}, {Wen}, {Briggs}, {Cline}, {Mazets}, {Golenetskii}, \&
  {Frederics}}]{slb+99}
{Smith}, D.~A. {\it et al.}, 1999, ApJ, 526, 683

\bibitem[{{Tavani}(1996)}]{tav96}
{Tavani}, M. 1996, ApJ, 466, 768

\bibitem[{{Vreeswijk} {et~al.}(2001){Vreeswijk}, {Fruchter}, {Kaper}, {Rol},
  {Galama}, {van Paradijs}, {Kouveliotou}, {Wijers}, {Pian}, {Palazzi},
  {Masetti}, {Frontera}, {Savaglio}, {Reinsch}, {Hessman}, {Beuermann},
  {Nicklas}, \& {van den Heuvel}}]{vfk+00}
{Vreeswijk}, P.~M. {\it et al.}, 2001, ApJ, 546, 672

\bibitem[{Vreeswijk {et~al.}(1999)Vreeswijk, Rol, Hjorth, Kouveliotou, Pian,
  Palazzi, Pedersen, Gorosabel, Castro-Tirado, \& Greiner}]{vrh+99}
Vreeswijk, P.~M. {\it et al.}, 1999, {GCN} notice 496

\bibitem[{{Woods} \& {Loeb}(1994)}]{wl94}
{Woods}, E. and {Loeb}, A. 1994, ApJ, 425, L63

\bibitem[{Woods {et~al.}(1998)Woods, Kippen, \& Connaughton}]{wkc98}
Woods, P., Kippen, R.~M., and Connaughton, V. 1998, {GCN} notice 112

\end{thebibliography}
\end{document}